\documentclass[prd,superscriptaddress,twocolumn,showpacs,showkeys,amsmath,amssymb]{revtex4}
\usepackage{amsmath}
\usepackage{graphicx}
\usepackage{color}
\usepackage{dcolumn}
\usepackage{bm}
\usepackage{slashed}

\newcommand{\bea}{\begin{eqnarray*}}
\newcommand{\eea}{\end{eqnarray*}}
\newcommand{\bean}{\begin{eqnarray}}
\newcommand{\eean}{\end{eqnarray}}
\newcommand{\eqs}[1]{Eqs.(\ref{#1})}
\newcommand{\eq}[1]{Eq.(\ref{#1})}
\newcommand{\meq}[1]{(\ref{#1})}
\newcommand{\grad}{\nabla}
\newcommand{\eqn}{&=&}
\newcommand{\non}{\nonumber \\}
\newcommand{\sgt}{\sqrt{h}}

\newcommand{\hsp}{\hspace{0.1mm}}

\newcommand{\pp}{\partial}

\begin{document}
\title{Consistency between dynamical and thermodynamical stabilities for perfect fluid in $f(R)$ theories}

\author{Xiongjun Fang}
\email[Xiongjun Fang: ]{fangxj@hunnu.edu.cn} \affiliation{Department of Physics, Key Laboratory of Low Dimensional Quantum Structures and
Quantum Control of Ministry of Education, and Synergetic Innovation Center for Quantum Effects and Applications, Hunan Normal University, Changsha, Hunan 410081, P. R. China}

\author{Xiaokai He}
\email[Xiaokai He: ]{hexiaokai77@163.com} \affiliation{Department of Physics, Key Laboratory of Low Dimensional Quantum Structures and Quantum Control of Ministry of Education, and Synergetic Innovation Center for Quantum Effects and Applications, Hunan Normal University, Changsha,
Hunan 410081, P. R. China}

\affiliation{School of Mathematics and Computational Science, Hunan First Normal University, Changsha 410205, China}

\author{Jiliang Jing}
\email[corresponding author: Jiliang Jing, ]{ jljing@hunnu.edu.cn} \affiliation{Department of Physics, Key Laboratory of Low Dimensional Quantum Structures and
Quantum Control of Ministry of Education, and Synergetic Innovation Center for Quantum Effects and Applications, Hunan Normal University, Changsha, Hunan 410081, P. R. China}

\begin{abstract}
We investigate the stability criterions for perfect fluid in $f(R)$ theories which is an important generalization of general relativity. Firstly, using Wald's general variation principle, we recast Seifert's work and obtain the dynamical stability criterion. Then using our generalized thermodynamical criterion, we obtain the concrete expressions of the criterion. We show that the dynamical stability criterion is exactly the same as the thermodynamical stability criterion. This result suggests that there is an inherent connection between the thermodynamics and gravity in $f(R)$  theories. It should be pointed out that using the thermodynamical method to determine the stability for perfect fluid is  simpler and more directly than the dynamical method. 
\end{abstract}


\maketitle

\section{Introduction}
The idea that there exist some deep connections between thermodynamics and gravity has been accepted widely since the establishment of the black hole thermodynamic laws. Some important works further reveal the relation between gravity and thermodynamics. Jacobson considered that the Einstein equation can be derived from thermodynamical relation which hold on Rindler causal horizons \cite{Jacobson1}, or the equilibrium of total entanglement entropy in ``Causal Diamond" \cite{Jacobson2}. Verlinder suggested that the gravity is the entropy force \cite{Verlinder}. In recent years, the proofs of the maximum entropy principle showed that the gravitational equation can be derived from the constraint equation and the maximum of total entropy \cite{gao,fang1,fang2,fang3,Cao1,Cao2}. All these researches are trying to establish a correspondence between the first variation of thermodynamic quantities and gravitational equation. However, assuming that the thermodynamic relation contains all information of gravity, it is naturally to investigate whether the second variation of thermodynamic quantities corresponds to the first variation of gravitational equations.

It is well-known that using the first variation of gravitational equation one can obtain the dynamical stability criterion. Chandrasekhar first discussed this problem and got the stability criterion for perfect fluid in general relativity \cite{Chandrasekhar1}. Based on the works of Chandrasekhar, Friedman and Schutz \cite{Friedman1,Friedman2,Friedman3}, Friedman defined ``canonical energy" and considered that it can provide the stability criterion \cite{Friedman4}. Seifert and Wald developed a general method to obtain the dynamical stability criterion for spherically symmetric perturbation in diffeomoephism covariant theories \cite{wald2007}. Meanwhile, the second variation of thermodynamic quantities, such as total entropy, can provide the thermodynamical stability criterion. A system is thermodynamical stable means that the system is in the thermodynamical equilibrium and the second variation of the total entropy of the system is negative, $\delta^2S<0$. Using thermodynamical method to handle the stability problem is more directly than dynamical method.

An interesting question is whether the dynamical method can be replaced by thermodynamical method. In other words, one can ask whether the dynamical stability criterion is the same as the thermodynamical stability criterion. In fact, Cocke presented the maximum entropy principle, and suggested that the thermodynamical stability is the same as dynamical stability \cite{Cocke}. Recently, Wald et al. \cite{wald2013} proved that in general relativity the thermodynamical stability is the same as the ``canonical energy" presented by Friedman. With the definition of AMD mass, the proofs in Ref. \cite{wald2013} need a crucial assumption that the spacetime should be asymptotically flat. Roupas \cite{Roupas} also proved that the maximum of total entropy for perfect fluid gives the same criterion for dynamical stability obtained by Yabushita \cite{Yabu}.

In Ref. \cite{fangsta}, we presented a generalized thermodynamical criterion, which is the second variation of total entropy for perfect fluid star. And we showed that in general relativity it can provide the same stability criterion as the dynamical stability criterion obtained by Wald \cite{wald2007}. It should be mentioned that all the previous works are focused on the cases in general relativity. However, whether the dynamical method can replaced by thermodynamical method in modified theories is not clear. As an important generalization of general relativity, $f(R)$ theories can explain the accelerated expansion of the universe because it contains higher order invariants in the action \cite{fR1, fR2}. In this manuscript, we show that the dynamical stability criterion is the same as the thermodynamical stability criterion in $f(R)$ theories, which implies that there is an inherent connection between thermodynamics and gravity.

The rest of this paper is organized as follows. In Section II, we briefly review wald's general variation principle. Since the result in Ref. \cite{Seifert} can not directly degenerate to general relativity, we recast the process of how to obtain the stability criterion by dynamical method. And our result can directly degenerate to general relativity. In Section III, we introduce the general thermodynamical stability criterion firstly and then show how to directly determine the stability criterion by thermodynamical method. At last, we summarized our manuscript with some discussions. It should be noted that only the basic idea and the main results are presented in Section II and Section III. The complicated derivation processes are described in Appendix B and Appendix C.

Throughout our manuscript, we use the sign conventions of Ref. \cite{waldbook}. Units will be those in which $c=G=1$, and the factor $\kappa=8\pi$ in gravitational equations will be ignored. We denote $f_R\equiv \frac{\pp f}{\pp R}$ and $b=\delta(f_R)=f_{RR}\delta R$. In order to distinguish $\mathcal{S}$ in Ref. \cite{Seifert} from the total entropy $S$ in our manuscript, we introduced a new quantity written as $\mathcal{F}=f_R'+\frac{2}{r}f_R$.

\section{Dynamical Method}
\subsection{Wald's general variation principle}
In this subsection we will briefly review the general variation principle presented by Seifert and Wald \cite{wald2007}. Referring to Ref. \cite{wald1994,wald1995} makes it available to obtain more details and discussions. Consider a diffeomorphism covariant Lagrangian four-form $\boldsymbol{\mathcal{L}}$, constructed from dynamical field $\Psi$, which consist of the spacetime metric $g_{ab}$ and other additional fields. The first variation of $\boldsymbol{\mathcal{L}}$ can be written as
\bean
\delta \boldsymbol{\mathcal{L}} = \boldsymbol{\mathcal{E}}\delta\Psi+d\boldsymbol{\theta} \,,
\eean
where $\boldsymbol{\mathcal{E}}=0$ defines the Euler-Lagrange equation of motion, and $\boldsymbol{\theta}$ is the symplectic potential three-form $\boldsymbol{\theta}(\Psi, \delta\Psi)$. The antisymmetrized variation of the symplectic potential $\boldsymbol{\theta}$ yields the symplectic current three-form $\boldsymbol{\omega}$ as
\bean \label{symcur}
\boldsymbol{\omega}=\delta_1\boldsymbol{\theta}(\Psi, \delta_2\Psi)-\delta_2\boldsymbol{\theta}(\Psi, \delta_1\Psi) \,.
\eean
When $\delta_1\Psi$ and $\delta_2\Psi$ satisfied the linearized equations of motions, then the symplectic current is conserved
\bean
d\boldsymbol{\omega}=0 \,.
\eean
For static background, the symplectic form $\boldsymbol{\Omega}$ for the theory can obtained by the integral of the pullback of symplectic current three-form
\bean
\boldsymbol{\Omega}(\Psi; \delta_1\Psi, \delta_2\Psi)=\int_{\Sigma}\boldsymbol{\bar\omega}[\Psi; \delta_1\Psi, \delta_2\Psi] \,,
\eean
where $\boldsymbol{\omega}$ is the pullback of $\boldsymbol{\omega}$ to the static hypersurfaces of the background solution.

If the perturbational fields are denoted as $\psi^{\alpha}$, then the symplectic form takes the form
\bean
\Omega(\Psi; \psi_1^{\alpha}, \psi_2^{\alpha})=\int_{\Sigma}\mathbf{W}_{\alpha\beta}
\left(\frac{\pp\psi_1^{\alpha}}{\pp t}\psi_2^{\beta}-\frac{\pp\psi_2^{\alpha}}{\pp t}\psi_1^{\beta}\right) \,,
\eean
where $\mathbf{W}_{\alpha\beta}$ is the three-form and it was showed that $\mathbf{W}_{\alpha\beta}=\mathbf{W}_{\beta\alpha}$ \cite{wald2007}.

Using $\mathbf{W}_{\alpha\beta}$ , we can define an inner product as
\bean \label{inner}
(\psi_1, \psi_2)\equiv \int_{\Sigma}\mathbf{W}_{\alpha\beta}\psi_1^{\alpha}\psi_2^{\beta}
\eean

Now suppose that the perturbational equations of motion take the form \cite{wald2007}
\bean \label{ppphippt}
-\frac{\pp^2}{\pp t^2}\psi^{\alpha}=\mathcal{T}^{\alpha}\hsp_{\beta}\psi^{\beta} \,,
\eean
where $\mathcal{T}$ is the time-evolution operator that contain spatial derivatives only. Seifert and Wald \cite{wald2007} showed that the operator $\mathcal{T}$ is self-adjoint and symmetric. Based on Rayleigh-Ritz principle, we know that the greatest lower bound, $\omega_0^2$, of the spectrum of $\mathcal{T}$ is
\bean \label{omega0}
\omega_0^2\leq \frac{(\psi, \mathcal{T}\psi)}{(\psi, \psi)} \,.
\eean
Wald \cite{wald1991} showed that the background solution is stable if $\omega_0^2>0$. The denominator of \eq{omega0} gives the inner product and always be positive. So the numerator $(\psi, \mathcal{T}\psi)>0$ implies that the system is stable, otherwise the perturbation exist a grow exponentially on a timescale $\tau=1/|\omega_0|$.

By \eqs{inner} and \meq{ppphippt}, the numerator of \eq{omega0} becomes
\bean \label{psiTpsi}
(\psi, \mathcal{T}\psi) = \int_{\Sigma}\mathbf{W}_{\alpha\beta}\psi_1^{\alpha}\mathcal{T}^{\beta}\hsp_{\gamma}\psi_2^{\gamma} = -\int_{\Sigma}\mathbf{W}_{\alpha\beta}\psi_1^{\alpha}\frac{\pp^2\psi_2^{\beta}}{\pp t^2} \,.
\eean
In next subsection, we give the dynamical stability criterion for $f(R)$ theories by \eq{psiTpsi}.

\subsection{Dynamical stability criterion for $f(R)$ theories}

\emph{\textbf {$ f(R)$ theories}}.~~  In $f(R)$ theories, the Ricci scalar $R$ in Einstein-Hilbert action is replaced by a function of $R$. The Lagrangian four-form $\boldsymbol{\mathcal{L}}$ reads as
\bean \label{lagrangianfR}
\boldsymbol{\mathcal{L}}=\frac{1}{2}f(R)\boldsymbol{\epsilon}+\boldsymbol{\mathcal{L}}_{matter}[A, g^{ab}] \,,
\eean
where $A$ denotes the matter fields. Taking the variation of this Lagrangian we obtain the equation of motion
\bean \label{fRgra}
f_R R_{ab}-\frac{1}{2}f g_{ab}-(\grad_a\grad_b-g_{ab}\Box)f_R=8\pi T_{ab} \,.
\eean
Similarly to \cite{Seifert}, $f(R)$ theories can be reduced to general relativity coupling to a scalar field $\alpha$. And the Lagrangian is given by
\bean
\boldsymbol{\mathcal{L}}=\frac{1}{2}(f_R(\alpha)R+f(\alpha)-\alpha f_R(\alpha))\epsilon
+\boldsymbol{\mathcal{L}}_{matter}[A, g^{ab}] \,.
\eean
Varying this Lagrangian yields
\bean \label{thetafR}
\theta^a_{f(R)} \eqn f_R\theta^a_{GR}+\theta^a_{matter} \non
\eqn +\frac{1}{2}[(\grad_b f_R)\delta g^{ab}-(\grad^af_R)g_{bc}\delta g^{bc}]  \,,
\eean
where $\theta_{GR}$ is the symplectic potential current in general relativity,
\bean
\theta_{GR}=\frac{1}{2}(g_{bc}\grad^a\delta g^{bc}-\grad_b\delta g^{ab}) \,.
\eean

\emph{\textbf {Spacetime metric.}} ~~ For spherically symmetric perturbations of static background, spherically symmetric spacetimes, the metric take the form \cite{MTW}
\bean \label{metric}
ds^2=-e^{2\Phi(t,r)}dt^2+e^{2\Lambda(t,r)}dr^2+r^2d\Omega^2 \,,
\eean
for some function $\Phi$ and $\Lambda$, where $d\Omega^2=d\theta^2+\sin^2\theta d\varphi^2$. In the rest of this manuscript, we denote the first order perturbation of $\Phi$ and $\Lambda$ by $\phi$ and $\lambda$, respectively, which means that $\Phi(t,r)=\Phi(r)+\phi(t,r)$ and $\Lambda(t,r)=\Lambda(r)+\lambda(t,r)$.

\emph{\textbf {Lagrangian for perfect fluid.}}~~ Following Ref.
\cite{wald2007}, we can define the three-form $N_{abc}$ on
four-dimensional manifold $M$, which represents the density of
fluid worldlines, by \bean N_{abc}=N_{ABC}(X)\grad_a X^A\grad_b
X^B\grad_c X^C \eean where $N_{ABC}$ is the three-form which was
defined on the three-dimensional manifold of fluid worldlines,
$\mathcal{M}$. Then one can define the scalar particle number
density $\nu$ in terms of $N_{abc}$ as \bean \label{nu2}
\nu^2=\frac{1}{6}N_{abc}N^{abc} \,. \eean If the Lagrangian for the matter part is given by \bean \label{Lmat}
\boldsymbol{\mathcal{L}}_{matter}=-\varrho(\nu)\boldsymbol{\epsilon} \,,
\eean where $\varrho$ is an arbitrary function of the particle
density $\nu$. Taking variation of \eq{Lmat}, Seifert and Wald
\cite{wald2007} constructed an identification \bean
\label{prhoidentify} \varrho \rightarrow \rho \,, \qquad
\varrho'\nu-\varrho \rightarrow p \,, \eean
Here $\varrho'$ denotes the derivative of $\varrho$ with respect to $\nu$. Except for this, $'$ means $\frac{\pp}{\pp r}$ in other situations in our manuscript. Moreover, they obtained the equation of motion of matter, which was just the relativistic Euler equation, and the symplectic
current $\boldsymbol\omega_{matter}$ of the perfect fluid.

From the conservation equation of energy momentum, $\grad_aT^{ab}=0$, we have
\bean
\frac{dp}{dr}=-(p+\rho)\frac{d\Phi}{dr} \,.
\eean
Together with the identification \eq{prhoidentify}, it is easy to find that
\bean \label{ppnuppr}
\frac{\pp\nu}{\pp r}=-\frac{\varrho'}{\varrho''}\frac{\pp\Phi}{\pp r}\,.
\eean
This formula would be used frequently in the subsequent calculations.

\emph{\textbf {Lagrangian displacement.}}~~ Seifert and Wald \cite{wald2007} used the ``Lagrangian coordinate" formalism to describe the fluid matter. In this formalism, the fluid is described by the manifold $\mathcal{M}$ of all fluid worldlines in the spacetime. There is a three ``fluid coordinate" $X^A$ on $\mathcal{M}$, $X^A\equiv \{X^R, X^{\Theta}, X^{\Phi}\}$, and in the static background solution we have $X^R=r$, $X^{\Theta}=\theta$, $X^{\Phi}=\varphi$. For spherically symmetric perturbation case, $\delta X^{\Theta}=\delta X^{\phi}=0$. The radial perturbation $\delta X^R$ can be describes by radial ``Lagrangian displacement" as
\bean
\xi(r,t)\equiv -\delta X^R(t,r) \,.
\eean
Note that we use the opposite signature of the definition of $\xi$ as Ref. \cite{wald2007}, to make the subsequent calculations appear self-consistent (See Appendix (A)).

\emph{\textbf {Dynamical stability criterion.}}~~ To get the dynamical stability criterion, we solve the linearized perturbation of gravitational equations firstly. The perturbation equation of $t-r$ and $r-r$ components of \eq{fRgra} gives
\bean \label{lambda}
\lambda=\mathcal{F}\left(\frac{\pp b}{\pp r}-\frac{\pp\Phi}{\pp r}b-e^{2\Lambda}\varrho'\nu\xi\right) \,,
\eean
and
\bean \label{bevo}
&& e^{2\Lambda-2\Phi}\frac{\pp^2b}{\pp t^2} = \mathcal{F}\frac{\pp\phi}{\pp r}
+\left(\frac{\pp\Phi}{\pp r}+\frac{2}{r}\right)\frac{\pp b}{\pp r} \non
&& ~\ +\left(\frac{2}{r}\frac{\pp\Lambda}{\pp r}+\frac{\pp\Lambda}{\pp r}\frac{\pp\Phi}{\pp r}-\left(\frac{\pp\Phi}{\pp r}\right)^2
-\frac{\pp^2\Phi}{\pp r^2}\right)b \non
&& ~\ -\left(2\left(\frac{\pp\Phi}{\pp r}+\frac{2}{r}\right)\mathcal{F}-\frac{6}{r^2}f_R\right)\lambda
-e^{2\Lambda}\varrho''\nu\delta\nu \,.
\eean
There is another conservation equation will be used. For spherical-symmetry metric \eq{metric}, the first order variation of scalar curvature can be written as
\bean \label{lambdaevo}
&& e^{2\Lambda}\left(R\lambda+\frac{b}{2f_{RR}}\right) = \frac{2e^{2\Lambda}}{r^2}\lambda-\frac{2}{r}\frac{\pp\phi}{\pp r}
-2\frac{\pp\Phi}{\pp r}\frac{\pp\phi}{\pp r} \non
&& +\frac{2}{r}\frac{\pp\lambda}{\pp r}+\frac{\pp\Phi}{\pp r}\frac{\pp\lambda}{\pp r}
+\frac{\pp\phi}{\pp r}\frac{\pp\Lambda}{\pp r}-\frac{\pp^2\phi}{\pp r^2}+e^{2\Lambda-2\Phi}\frac{\pp^2\lambda}{\pp t^2} \,.
\eean
While we also have the matter equation of motion \cite{wald2007}, which is just the time-evolution equation of variable $\xi$
\bean \label{xievo}
e^{2\Lambda-2\Phi}\varrho'\frac{\pp^2\xi}{\pp t^2}+\varrho'\frac{\pp\phi}{\pp r}
+\left(\frac{\pp}{\pp r}+\frac{\pp\Phi}{\pp r}\right)(\varrho''\delta\nu)=0 \,.
\eean

Substituting \eq{thetafR} into \eq{symcur}, we find the specific form of the t-component of the symplectic current in $f(R)$ theories takes the form \cite{Seifert}:
\bean \label{omegat}
&& \omega^t = \varrho'\nu e^{2\Lambda-2\Phi}\left(\frac{\pp\xi_1}{\pp t}\xi_2-\frac{\pp\xi_2}{\pp t}\xi_1\right) \non
&& ~\ +e^{-2\Phi}\left(b_1\frac{\pp\lambda_2}{\pp t}-\frac{\pp b_1}{\pp t}\lambda_2
-b_2\frac{\pp\lambda_1}{\pp t}+\frac{\pp b_2}{\pp t}\lambda_1\right) \,.
\eean
From \eq{omegat} we can read off $\mathbf{W}_{\alpha\beta}$. Denote the numerator part of $\omega_0$ in \eq{omega0}, $(\psi, \mathcal{T}\psi)$,  by $\mathcal{P}_{Dyn}$. More explicitly, together with \eqs{lambdaevo} and \meq{xievo}, it can be written as
\bean \label{PDyn}
\mathcal{P}_{Dyn} \eqn  \int_r -r^2e^{3\Lambda-\Phi}\varrho'\nu\xi\frac{\pp^2\xi}{\pp t^2} \non
&& +r^2 e^{\Lambda-\Phi}\lambda\frac{\pp^2b}{\pp t^2}+r^2e^{\Lambda-\Phi}b\frac{\pp^2\lambda}{\pp t^2} \,.
\eean
To obtain the time-evolution of each variables, i.e. $\frac{\pp^2\xi}{\pp t^2}$, $\frac{\pp^2 b}{\pp t^2}$ and $\frac{\pp^2\lambda}{\pp t^2}$, Seifert \cite{Seifert} showed that one can derive $\pp\phi/\pp r$ from the isotropy condition for perfect fluid, $p=G_r\hsp^r=G_{\theta}\hsp^{\theta}$, to eliminate the $\Box f_R$ terms. However, the expression of $\pp\phi/\pp r$ derived from Ref. \cite{Seifert} can not directly degenerate to general relativity case. Here we adopt some different techniques without any other constraint to handle \eq{PDyn}, and find that our result can degenerate to general relativity. Using integration by parts and after some complicated calculations, we finally obtain the dynamical stability criterion in the form
\bean \label{DynResult}
\mathcal{P}_{Dyn} \eqn \int_r \mathcal{C}_{Dyn}^1\left(\frac{\pp b}{\pp r}\right)^2+\mathcal{C}_{Dyn}^2\left(\frac{\pp\xi}{\pp r}\right)^2
+\mathcal{C}_{Dyn}^3\xi\frac{\pp b}{\pp r} \non
&& +\mathcal{C}_{Dyn}^4b\frac{\pp\xi}{\pp r}+\mathcal{C}_{Dyn}^5\frac{\pp b}{\pp r}\frac{\pp \xi}{\pp r}+\mathcal{C}_{Dyn}^6b^2 \non
&& +\mathcal{C}_{Dyn}^7\xi^2+\mathcal{C}_{Dyn}^8b\xi \,,
\eean
where each $\mathcal{C}_{Dyn}^i$ are the functions of the background fields. We put all detailed calculations and the expressions of each $\mathcal{C}_{Dyn}^i$ in Appendix (B).

\section{Thermodynamical Method}
We first briefly introduce the generalized thermodynamical stability criterion obtained from the maximum entropy principle. Please refer to Refs. \cite{fang1} and \cite{fangsta} to obtain further details and discussion.

Considering a perfect fluid star in static background spacetime with a static slice $\Sigma$, we assume that quantities are measured by static observers. The Tolman's law gives $T\chi=const.$, where $T$ and $\chi$ are the temperature of the fluid and the redshift factor, respectively. Without loss of generality, we take $T\chi=1$. Assuming that the fluid in region $C$ on $\Sigma$ satisfied ordinary thermodynamic relations and Tolman's law. We have shown that if the constraint equation holds and the total entropy is an extremum, then the other components of gravitational equation are obtained \cite{fang1,fang3,Cao2}. The fact implies that the full set of gravitational equations may be replaced by the ordinary thermodynamic relations and the constraint equation.

An isolated system in thermodynamical equilibrium is said to be thermodynamical stable if $\delta^2S<0$. In Ref. \cite{fangsta} we have shown that if the star is deviated slightly from equilibrium state, the second variation of total entropy takes the form
\bean \label{delta2S}
\delta^2S \eqn \int_C\frac{1}{T}\left[2\delta\rho\delta\sgt+\sgt\delta^2\rho \right. \non
&& \left. +(p+\rho)\delta^2\sgt-\frac{\delta p \cdot \delta\rho}{p+\rho}\sgt\right] \,,
\eean
where $\rho$ and $p$ are the energy density and pressure, respectively.

In the case of general relativity, we have proven that the thermodynamical stability criterion given by \eq{delta2S} is the same as the dynamical stability criterion given by Ref. \cite{wald2007}. In this section, we calculate the thermodynamical stability criterion in $f(R)$ gravities. Recall that the gravitational field equation of $f(R)$ theories can be written as
\bean \label{fRgrathermo}
f_RR_{ab}-\frac{1}{2}f g_{ab}-[\grad_a\grad_b-g_{ab}\Box]f_R=8\pi T_{ab} \,,
\eean
and the general spherical symmetric spacetime is
\bean
ds^2=-e^{2\Phi(t,r)}dt^2+e^{2\Lambda(t,r)}dr^2+r^2d\Omega^2 \,,
\eean
we can easily obtain the energy density $\rho=T_t\hsp^t$ and the pressure $p=T_r\hsp^r$ of the perfect fluid star as
\bean \label{fRrho}
\rho \eqn \frac{f_R}{r^2}+\frac{e^{-2\Lambda}(2r\Lambda'-1)} {r^2}f_R+\frac{1}{2}f \non
&& -\frac{R}{2}f_R-e^{-2\Lambda}f_R'' +e^{-2\Lambda}\left(\Lambda'-\frac{2}{r}\right)f_R' \,,
\eean
\bean \label{fRp}
p \eqn -\frac{f_R}{r^2}+\frac{e^{-2\Lambda}(1+2r\Phi')} {r^2}f_R-\frac{1}{2}f \non
&& +\frac{R}{2}f_R+e^{-2\Lambda}\left(\frac{2}{r}+\Phi'\right)f_R' \,.
\eean
Since the redshift factor $\chi=e^{\Phi}$, the temperature takes the form $T=e^{-\Phi}$. Based on Chandrasekhar's procedure \cite{Chandrasekhar1}, the $t-r$ component of \eq{fRgrathermo} gives
\bean \label{G01T01fR}
-(p+\rho)e^{2\Lambda}\frac{\pp\xi}{\pp x^0} \eqn
f_R\frac{2}{r}\frac{\pp\Lambda}{\pp x^0}-\frac{\pp f_R'}{\pp x^0} \non
&& +\frac{\pp\Phi}{\pp r}\frac{\pp f_R}{\pp x^0}
+\frac{\pp\Lambda}{\pp x^0}\frac{\pp f_R}{\pp r} \,.
\eean
By direct integration, we have
\bean \label{Thermolambda}
\lambda = \mathcal{F}^{-1}\left(\frac{\pp}{\pp r}b-\frac{\pp\Phi}{\pp r}b-(p+\rho)e^{2\Lambda}\xi\right) \,,
\eean
this is just \eq{lambda}.

To calculate the explicitly form of \eq{delta2S} and compare with the dynamical stability criterion, we consider the same Lagrangian of matter as \eq{Lmat}, which gives the identified relation \eq{prhoidentify} as
\bean
\varrho \rightarrow \rho \,, \qquad \varrho'\nu-\varrho \rightarrow p \,.
\eean
These relations yield
\bean
\delta\rho=\varrho'\delta\nu \,, \qquad \delta p=\varrho''\nu\cdot\delta\nu \,,
\eean
so
\bean
\frac{\delta p \cdot \delta\rho}{p+\rho}=\frac{\varrho''\nu\delta\nu\cdot \varrho'\delta\nu}{\varrho'\nu}=\varrho''(\delta\nu)^2 \,,
\eean
where the explicitly form of $\delta\nu$ is given by \eq{delnu}. Meanwhile, the variation of the volume element $\sgt$ gives
\bean \label{delsgt}
\delta\sgt=r^2e^{\Lambda}\lambda \,,
\eean
and the variation of \eq{fRrho} gives
\bean \label{firstvariationrho}
\delta\rho \eqn \frac{b}{r^2}-\frac{2e^{-2\Lambda}}{r^2}\left(2r\frac{\pp\Lambda}{\pp r}-1\right)f_R\lambda
+\frac{2e^{-2\Lambda}}{r}f_R \frac{\pp\lambda}{\pp r} \non
&& +\frac{e^{-2\Lambda}}{r^2}\left(2r\frac{\pp\Lambda}{\pp r}-1\right)b-\frac{R}{2}b+2e^{-2\Lambda}f_R''\lambda \non
&& -e^{-2\Lambda}b''-2e^{-2\Lambda}\left(\frac{\pp\Lambda}{\pp r}-\frac{2}{r}\right)f_R'\lambda \non
&& +e^{-2\Lambda}\frac{\pp\lambda}{\pp r}f_R'+e^{-2\Lambda}\left(\frac{\pp\Lambda}{\pp r}-\frac{2}{r}\right)b' \,.
\eean
Then the second variation of $\sgt$ and $\rho$ can be obtained. Substituting these relations into \eq{delta2S}, after tediously calculations we find that the thermodynamical stability criterion takes the form:
\bean \label{ThermoResult}
\delta^2S \eqn \int_r \mathcal{C}_{Thermo}^1\left(\frac{\pp b}{\pp r}\right)^2+\mathcal{C}_{Thermo}^2\left(\frac{\pp\xi}{\pp r}\right)^2 \non
&& +\mathcal{C}_{Thermo}^3\xi\frac{\pp b}{\pp r}+\mathcal{C}_{Thermo}^4b\frac{\pp\xi}{\pp r}
+\mathcal{C}_{Thermo}^5\frac{\pp b}{\pp r}\frac{\pp \xi}{\pp r} \non
&& +\mathcal{C}_{Thermo}^6b^2+\mathcal{C}_{Thermo}^7\xi^2+\mathcal{C}_{Thermo}^8b\xi \,,
\eean
where each $\mathcal{C}_{Thermo}^i$ are the functions of the background fields. We put all detailed calculations and the expressions of each $\mathcal{C}_{Thermo}^i$ in Appendix (C).

\section{Conclusions and Discussions}

We have investigated the stability criterions for static spherical symmetric perfect fluid in $f(R)$ theories by dynamical and thermodynamical methods, respectively. The dynamical stability criterion of the spherically symmetric perfect fluid in $f(R)$ theories is given by \eq{DynResult}. If the system is dynamical stable, then $(\psi, \mathcal{T}\psi)>0$, i.e., $\mathcal{P}_{Dyn}>0$. And the thermodynamical stability criterion is given by \eq{ThermoResult}. The negative of the second variation of the total entropy, $\delta^2S<0$, means the system is thermodynamical stable.
Comparing \eqs{PDynC1}, \meq{PDynC2}, \meq{PDynC3}, \meq{PDynC4}, \meq{PDynC5}, \meq{PDynC6}, \meq{PDynC7}, \meq{PDynC8} and \eqs{STherC1}, \meq{STherC2}, \meq{STherC3}, \meq{STherC4}, \meq{STherC5}, \meq{STherC6}, \meq{STherC7}, \meq{STherC8}, we find that
\bean
\mathcal{C}_{Dyn}^i=-\mathcal{C}_{Thermo}^i \,.
\eean
Therefore,  Eqs. (\ref{DynResult}) and (\ref{ThermoResult}) are exactly the same except the opposite signs, which shows that the dynamical stability criterion is the same as the thermodynamical stability criterion in $f(R)$ theories.  Combining with the result obtained in Ref. \cite{fangsta},  we find that the dynamical stability can be replaced by thermodynamical stability not only in general relativity, but also in modified theories, such as $f(R)$ theories. It suggests that there is a universal inherent connection between thermodynamics and gravity.

Comparing the concrete calculations of the stability criterions, we found that thermodynamical method is more directly and simpler than the dynamical method. In dynamical method, one should determine the symplectic current $\omega^a$ and the three-form $\mathbf{W}_{\alpha\beta}$ firstly, and then solve each evolution equation for variable by complicated calculations (it should be pointed out whether the evolution equation can be solved is uncertainly). After these preparations one can obtain the stability criterion.  In thermodynamical method, we only need to substitute the expressions of the energy density $\rho$ and the pressure $p$ into \eq{delta2S}, then the thermodynamical stability criterion can be obtained directly.

\acknowledgments
Jing was supported by the NSFC (No.~11475061). He was supported by the NSFC (No.~11401199).

\appendix
\section{Signature of the definition of $\xi$}
In this appendix, we would show that if we choose an opposite signature of the definition of $\xi$ given by Ref. \cite{wald2007}, i.e.,
\bean
\xi=-\delta X^R \,,
\eean
then it seems self-consistent that the variation of energy density obtained by different ways would be the same.
Similarly to \cite{wald2007}, assume that
\bean
N_{R\Theta\Phi}=q(X^R)\sin X^{\Theta} \,,
\eean
where $q$ is an arbitrary function of $X^R$. By \eq{nu2}, the number density can be expressed as
\bean \label{nudef}
\nu=\frac{q(X^R)}{r^2}\sqrt{e^{-2\Lambda}\left(\frac{\pp X^R}{\pp r}\right)^2-e^{-2\Phi}\left(\frac{\pp X^R}{\pp t}\right)^2} \,.
\eean
If we choose $\xi=-\delta X^R$, then the variation of \eq{nudef} gives
\bean \label{delnu}
\delta\nu=-\nu\left(\frac{\pp\xi}{\pp r}+\left(\frac{1}{\nu}\frac{\pp\nu}{\pp r}+\frac{\pp\Lambda}{\pp r}+\frac{2}{r}\right)\xi+\lambda\right)\,.
\eean

In general relativity, following the main processes presented in \cite{wald2007}, we find that
\bean \label{lambdaGR}
\lambda_{GR}=-\frac{r}{2}e^{2\Lambda}\varrho'\nu\xi \,.
\eean
Together with the background equation of motion
\bean \label{varrhonu}
\varrho'\nu=p+\rho=\frac{2}{r}e^{-2\Lambda}\left(\frac{\pp\Lambda}{\pp r}+\frac{\pp\Phi}{\pp r}\right) \,,
\eean
substituting \eq{lambdaGR} into \eq{delnu} yields
\bean
\delta\nu_{GR} = -\nu\left(\frac{\pp\xi}{\pp r}+\frac{2}{r}\xi-\frac{\pp\Phi}{\pp r}\xi+\frac{1}{\nu}\frac{\pp\nu}{\pp r}\xi\right) \,.
\eean
However, the variation of the energy density $\delta\rho$ can also be written as \cite{Chandrasekhar2}
\bean
\delta\rho_{GR}=-\frac{1}{r^2}\frac{\pp}{\pp r}[r^2(p+\rho)\xi] \,,
\eean
combining with \eq{varrhonu}, directly calculation shows that
\bean \label{rhoGR}
\delta\rho_{GR} =\varrho'\delta\nu_{GR} \,.
\eean

Generalize to $f(R)$ theories, \eq{lambdaGR} becomes to
\bean
\lambda_{f(R)}=\mathcal{F}^{-1}\left(\frac{\pp b}{\pp r}-\frac{\pp\Phi}{\pp r}b-e^{2\Lambda}\varrho'\nu\xi\right) \,,
\eean
then
\begin{widetext}
\bean
\delta\nu_{f(R)}
\eqn -\nu\left(\frac{\pp\xi}{\pp r}+\left(\frac{1}{\nu}\frac{\pp\nu}{\pp r}+\frac{2}{r}+\frac{\pp\Lambda}{\pp r}\right)\xi+\lambda\right) \non
\eqn -\nu\left(\frac{\pp\xi}{\pp r}+\left(\frac{1}{\nu}\frac{\pp\nu}{\pp r}+\frac{2}{r}-\frac{\pp\Phi}{\pp r}\right)\xi
+\mathcal{F}^{-1}\frac{\pp b}{\pp r}-\mathcal{F}^{-1}\frac{\pp\Phi}{\pp r}b+f_R''\mathcal{F}^{-1} \right) \,.
\eean
Take the variation of \eq{fRrho} in $f(R)$ theories,
\bean \label{rhofR}
\delta\rho_{f(R)} \eqn -\varrho'\nu\frac{\pp\xi}{\pp r}-\mathcal{F}^{-1}\varrho'\nu\frac{\pp b}{\pp r}+\left[\varrho'\nu\frac{\pp\Phi}{\pp r}
+\frac{\varrho'^2}{\varrho''}\frac{\pp\Phi}{\pp r}-\frac{2}{r}\varrho'\nu-\varrho'\nu f_R''\mathcal{F}^{-1}\right]\xi
+\mathcal{F}^{-1}\varrho'\nu\frac{\pp\Phi}{\pp r}b \non
\eqn -\varrho'\nu\left[\frac{\pp\xi}{\pp r}+\mathcal{F}^{-1}\frac{\pp b}{\pp r}-\frac{\pp\Phi}{\pp r}\xi
+\frac{1}{\nu}\frac{\pp\nu}{\pp r}\xi+\frac{2}{r}\xi+f_R''\mathcal{F}^{-1}\xi-\mathcal{F}^{-1}\frac{\pp\Phi}{\pp r}b \right] \non
\eqn \varrho'\delta\nu_{f(R)} \,.
\eean
So \eq{rhoGR} and \eq{rhofR} show that it is self-consistent if we define the ``Lagrangian displacement" as $\xi=-\delta X^R$.

\section{Detailed calculation and result of Dynamical stability criterion}
In this appendix, we will show the detailed calculations of how to obtain the criterion for dynamical stability. The main goal is to eliminate the time-evolution terms in \eq{PDyn}. Some relationships are repeatedly used, such as \eq{ppnuppr}. We also use the background equation of motion, which takes the form
\bean \label{pplusrho}
p+\rho=\varrho'\nu=e^{-2\Lambda}\left(\frac{\pp\Phi}{\pp r}+\frac{\pp\Lambda}{\pp r}\right)\mathcal{F}-e^{-2\Lambda}f_R'' \,.
\eean

We already have time-evolution equations of variables $b$, $\lambda$ and $\xi$, see \eqs{bevo}, \meq{lambdaevo} and \meq{xievo}. From \eq{bevo}, we obtain
\bean \label{ppphi2ppr}
\mathcal{F}\frac{\pp^2\phi}{\pp r^2} \eqn -\frac{\pp\mathcal{F}}{\pp r}\frac{\pp\phi}{\pp r}
+2e^{2\Lambda-2\Phi}\left(\frac{\pp\Lambda}{\pp r}-\frac{\pp\Phi}{\pp r}\right)\frac{\pp^2b}{\pp t^2}
+e^{2\Lambda-2\Phi}\frac{\pp}{\pp r}\frac{\pp^2b}{\pp t^2}
-\left(\frac{\pp^2\Phi}{\pp r^2}-\frac{2}{r^2}\right)\frac{\pp b}{\pp r}
-\left(\frac{\pp\Phi}{\pp r}+\frac{2}{r}\right)\frac{\pp^2b}{\pp r^2} \non
&& -\left(-\frac{2}{r^2}\frac{\pp\Lambda}{\pp r}+\frac{2}{r}\frac{\pp^2\Lambda}{\pp r^2}
+\frac{\pp^2\Lambda}{\pp r^2}\frac{\pp\Phi}{\pp r}+\frac{\pp\Lambda}{\pp r}\frac{\pp^2\Phi}{\pp r^2}
-2\frac{\pp\Phi}{\pp r}\frac{\pp^2\Phi}{\pp r^2}-\frac{\pp^3\Phi}{\pp r^3}\right)b \non
&& -\left(\frac{2}{r}\frac{\pp\Lambda}{\pp r}+\frac{\pp\Lambda}{\pp r}\frac{\pp\Phi}{\pp r}-\left(\frac{\pp\Phi}{\pp r}\right)^2
-\frac{\pp^2\Phi}{\pp r^2}\right)\frac{\pp b}{\pp r} \non
&& +\left(2\left(\frac{\pp^2\Phi}{\pp r^2}-\frac{2}{r^2}\right)\mathcal{F}
+2\left(\frac{\pp\Phi}{\pp r}+\frac{2}{r}\right)\frac{\pp \mathcal{F}}{\pp r}
+\frac{12}{r^3}f_R-\frac{6}{r^2}f_R'\right)\lambda
+\left(2\left(\frac{\pp\Phi}{\pp r}+\frac{2}{r}\right)\mathcal{F}-\frac{6}{r^2}f_R\right)\frac{\pp\lambda}{\pp r} \non
&& +2e^{2\Lambda}\frac{\pp\Lambda}{\pp r}\varrho''\nu\delta\nu+e^{2\Lambda}\varrho'''\frac{\pp\nu}{\pp r}\nu\delta\nu
+e^{2\Lambda}\varrho''\frac{\pp\nu}{\pp r}\delta\nu+e^{2\Lambda}\varrho''\nu\frac{\pp\delta\nu}{\pp r} \,.
\eean
Note that \eq{lambda} yields
\bean \label{lambda2}
\mathcal{F}e^{2\Lambda-2\Phi}\frac{\pp^2\lambda}{\pp t^2} = e^{2\Lambda-2\Phi}\left(\frac{\pp}{\pp r}\frac{\pp^2b}{\pp t^2}
-\frac{\pp\Phi}{\pp r}\frac{\pp^2b}{\pp t^2}-e^{2\Lambda}\varrho'\nu\frac{\pp^2\xi}{\pp t^2}\right) \,.
\eean
Together with \eqs{lambdaevo} and \meq{lambda2}, and then using \eq{ppphi2ppr}, we obtain
\bean
&& \mathcal{F}e^{2\Lambda}\left(R-\frac{2}{r^2}\right)\lambda+\mathcal{F}\frac{e^{2\Lambda}b}{2f_{RR}}
+\mathcal{F}\left(\frac{2}{r}+2\frac{\pp\Phi}{\pp r}-\frac{\pp\Lambda}{\pp r}\right)\frac{\pp\phi}{\pp r}
-\mathcal{F}\left(\frac{\pp\Phi}{\pp r}+\frac{2}{r}\right)\frac{\pp\lambda}{\pp r}
+2e^{2\Lambda-2\Phi}\left(\frac{\pp\Lambda}{\pp r}-\frac{\pp\Phi}{\pp r}\right)\frac{\pp^2b}{\pp t^2} \non
&& -\left(\frac{\pp^2\Phi}{\pp r^2}-\frac{2}{r^2}\right)\frac{\pp b}{\pp r}
-\left(\frac{\pp\Phi}{\pp r}+\frac{2}{r}\right)\frac{\pp^2b}{\pp r^2}
-\left(-\frac{2}{r^2}\frac{\pp\Lambda}{\pp r}+\frac{2}{r}\frac{\pp^2\Lambda}{\pp r^2}
+\frac{\pp^2\Lambda}{\pp r^2}\frac{\pp\Phi}{\pp r}+\frac{\pp\Lambda}{\pp r}\frac{\pp^2\Phi}{\pp r^2}
-2\frac{\pp\Phi}{\pp r}\frac{\pp^2\Phi}{\pp r^2}-\frac{\pp^3\Phi}{\pp r^3}\right)b \non
&& -\left(\frac{2}{r}\frac{\pp\Lambda}{\pp r}+\frac{\pp\Lambda}{\pp r}\frac{\pp\Phi}{\pp r}-\left(\frac{\pp\Phi}{\pp r}\right)^2
-\frac{\pp^2\Phi}{\pp r^2}\right)\frac{\pp b}{\pp r}
+\left(2\left(\frac{\pp^2\Phi}{\pp r^2}-\frac{2}{r^2}\right)\mathcal{F}
+2\left(\frac{\pp\Phi}{\pp r}+\frac{2}{r}\right)\frac{\pp \mathcal{F}}{\pp r}
+\frac{12}{r^3}f_R-\frac{6}{r^2}f_R'\right)\lambda \non
&& +\left(2\left(\frac{\pp\Phi}{\pp r}+\frac{2}{r}\right)\mathcal{F}-\frac{6}{r^2}f_R\right)\frac{\pp\lambda}{\pp r}
+2e^{2\Lambda}\frac{\pp\Lambda}{\pp r}\varrho''\nu\delta\nu+e^{2\Lambda}\varrho'''\frac{\pp\nu}{\pp r}\nu\delta\nu
+e^{2\Lambda}\varrho''\frac{\pp\nu}{\pp r}\delta\nu+e^{2\Lambda}\varrho''\nu\frac{\pp\delta\nu}{\pp r}
-\frac{\pp\mathcal{F}}{\pp r}\frac{\pp\phi}{\pp r} \non
&& +e^{2\Lambda-2\Phi}\frac{\pp\Phi}{\pp r}\frac{\pp^2b}{\pp t^2}
-e^{2\Lambda}\varrho'\nu\frac{\pp\phi}{\pp r}-e^{2\Lambda}\nu\left(\frac{\pp}{\pp r}+\frac{\pp\Phi}{\pp r}\right)(\varrho''\delta\nu)
= 0 \,.
\eean
Simplified this relation we have
\bean \label{6bevo}
&& \frac{6}{r^2}\mathcal{F}^{-1}f_Re^{2\Lambda-2\Phi}\frac{\pp^2b}{\pp t^2}
= -\mathcal{F}e^{2\Lambda}\left(R-\frac{2}{r^2}\right)\lambda-\mathcal{F}\frac{e^{2\Lambda}b}{2f_{RR}}
+\left(\frac{\pp\Phi}{\pp r}+\frac{2}{r}\right)\frac{\pp^2b}{\pp r^2} \non
&& +\left(-\frac{2}{r^2}\frac{\pp\Lambda}{\pp r}+\frac{2}{r}\frac{\pp^2\Lambda}{\pp r^2}
+\frac{\pp^2\Lambda}{\pp r^2}\frac{\pp\Phi}{\pp r}+\frac{\pp\Lambda}{\pp r}\frac{\pp^2\Phi}{\pp r^2}
-2\frac{\pp\Phi}{\pp r}\frac{\pp^2\Phi}{\pp r^2}-\frac{\pp^3\Phi}{\pp r^3}\right)b \non
&& +\left(\frac{2}{r}\frac{\pp\Lambda}{\pp r}+\frac{\pp\Lambda}{\pp r}\frac{\pp\Phi}{\pp r}-\left(\frac{\pp\Phi}{\pp r}\right)^2
-\frac{2}{r^2}\right)\frac{\pp b}{\pp r} \non
&& -\left(2\left(\frac{\pp^2\Phi}{\pp r^2}-\frac{2}{r^2}\right)\mathcal{F}
+2\left(\frac{\pp\Phi}{\pp r}+\frac{2}{r}\right)\frac{\pp \mathcal{F}}{\pp r}
+\frac{12}{r^3}f_R-\frac{6}{r^2}f_R'\right)\lambda \non
&& +\left(\left(\frac{\pp\Phi}{\pp r}+\frac{2}{r}\right)-\frac{6}{r^2}\mathcal{F}^{-1}f_R\right)
\frac{\pp\mathcal{F}}{\pp r}\lambda-\left(\left(\frac{\pp\Phi}{\pp r}+\frac{2}{r}\right)-\frac{6}{r^2}\mathcal{F}^{-1}f_R\right) \non
&& \times \left(\frac{\pp^2b}{\pp r^2}-\frac{\pp^2\Phi}{\pp r^2}b-\frac{\pp\Phi}{\pp r}\frac{\pp b}{\pp r}
-2e^{2\Lambda}\frac{\pp\Lambda}{\pp r}\varrho'\nu\xi-e^{2\Lambda}\varrho''\frac{\pp\nu}{\pp r}\nu\xi
-e^{2\Lambda}\varrho'\frac{\pp\nu}{\pp r}\xi-e^{2\Lambda}\varrho'\nu\frac{\pp\xi}{\pp r}\right) \non
&& -e^{2\Lambda}\varrho''\frac{\pp\nu}{\pp r}\delta\nu
+\left(\frac{\pp\Phi}{\pp r}-2\frac{\pp\Lambda}{\pp r}+\frac{6}{r^2}\mathcal{F}^{-1}f_R\right)
\left(\frac{\pp\Phi}{\pp r}+\frac{2}{r}\right)\frac{\pp b}{\pp r} \non
&& +\left(\frac{\pp\Phi}{\pp r}-2\frac{\pp\Lambda}{\pp r}+\frac{6}{r^2}\mathcal{F}^{-1}f_R\right)\left(\frac{2}{r}\frac{\pp\Lambda}{\pp r}
+\frac{\pp\Lambda}{\pp r}\frac{\pp\Phi}{\pp r}-\left(\frac{\pp\Phi}{\pp r}\right)^2-\frac{\pp^2\Phi}{\pp r^2}\right)b \non
&& -\left(\frac{\pp\Phi}{\pp r}-2\frac{\pp\Lambda}{\pp r}+\frac{6}{r^2}\mathcal{F}^{-1}f_R\right)
\left(2\left(\frac{\pp\Phi}{\pp r}+\frac{2}{r}\right)\mathcal{F}-\frac{6}{r^2}f_R\right)\lambda
-\frac{6}{r^2}\mathcal{F}^{-1}f_Re^{2\Lambda}\varrho''\nu\delta\nu \,.
\eean

Now we pay attention to \eq{PDyn}
\bean \label{B7}
\mathcal{P}_{Dyn} = \int_r -r^2e^{3\Lambda-\Phi}\varrho'\nu\xi\frac{\pp^2\xi}{\pp t^2}
+r^2 e^{\Lambda-\Phi}\lambda\frac{\pp^2b}{\pp t^2}+r^2e^{\Lambda-\Phi}b\frac{\pp^2\lambda}{\pp t^2} \,.
\eean
Substituting \eqs{lambdaevo} and \meq{xievo} into \eq{B7}, and using integration by parts, then
\bean
\mathcal{P}_{Dyn} \eqn \int_r r^2e^{\Lambda+\Phi}\xi\varrho'\nu\frac{\pp\phi}{\pp r}
-2re^{\Lambda+\Phi}\nu\varrho''\xi\delta\nu
-r^2e^{\Lambda+\Phi}\frac{\pp\Lambda}{\pp r}\nu\varrho''\xi\delta\nu
-r^2e^{\Lambda+\Phi}\nu\varrho''\frac{\pp\xi}{\pp r}\delta\nu \non
&& +r^2e^{\Lambda+\Phi}\varrho'\frac{\pp\Phi}{\pp r}\xi\delta\nu
+r^2 e^{\Lambda-\Phi}\mathcal{F}^{-2}\left(-e^{2\Lambda}\varrho'\nu b-\frac{6}{r^2}f_R b
-\mathcal{F}e^{2\Lambda}\varrho'\nu\xi\right)\frac{\pp^2b}{\pp t^2} \non
&& +r^2e^{\Lambda+\Phi}b\mathcal{F}^{-1}\varrho'\nu\frac{\pp\phi}{\pp r}
-2re^{\Lambda+\Phi}\mathcal{F}^{-1}\varrho''\nu b\delta\nu
-r^2e^{\Lambda+\Phi}\frac{\pp\Lambda}{\pp r}\mathcal{F}^{-1}\varrho''\nu b\delta\nu \non
&& +r^2e^{\Lambda+\Phi}\frac{\pp\Phi}{\pp r}\mathcal{F}^{-1}\varrho'b\delta\nu
-r^2e^{\Lambda+\Phi}\mathcal{F}^{-1}\varrho''\nu\frac{\pp b}{\pp r}\delta\nu
+r^2\mathcal{F}^{-2}e^{\Lambda+\Phi}\frac{\pp \mathcal{F}}{\pp r}\varrho''\nu b\delta\nu \,.
\eean
Together with \eqs{bevo} and \meq{6bevo}, all time-evolution terms eliminated in $\mathcal{P}_{Dyn}$,
\bean \label{PDynFinal}
\mathcal{P}_{Dyn}
\eqn \int_r r^2e^{\Lambda+\Phi}\mathcal{F}^{-1}\left\{ -\varrho'\nu\left(\frac{\pp\Phi}{\pp r}+\frac{2}{r}\right)\xi\frac{\pp b}{\pp r}
-\varrho'\nu\left(\frac{2}{r}\frac{\pp\Lambda}{\pp r}
+\frac{\pp\Lambda}{\pp r}\frac{\pp\Phi}{\pp r}-\left(\frac{\pp\Phi}{\pp r}\right)^2
-\frac{\pp^2\Phi}{\pp r^2}\right)\xi b \right. \non
&& +\varrho'\nu\left(2\left(\frac{\pp\Phi}{\pp r}+\frac{2}{r}\right)\mathcal{F} -\frac{6}{r^2}f_R\right)\xi\lambda
+\varrho'\nu e^{2\Lambda}\varrho''\nu\xi\delta\nu
-\frac{2}{r}\mathcal{F}\nu\varrho''\xi\delta\nu
-\mathcal{F}\frac{\pp\Lambda}{\pp r}\nu\varrho''\xi\delta\nu \non
&& -\mathcal{F}\nu\varrho''\frac{\pp\xi}{\pp r}\delta\nu
+\mathcal{F}\varrho'\frac{\pp\Phi}{\pp r}\xi\delta\nu
-\mathcal{F}^{-1}\varrho'\nu\left(\frac{\pp\Phi}{\pp r}+\frac{2}{r}\right)b\frac{\pp b}{\pp r} \non
&& -\mathcal{F}^{-1}\varrho'\nu\left(\frac{2}{r}\frac{\pp\Lambda}{\pp r}+\frac{\pp\Lambda}{\pp r}\frac{\pp\Phi}{\pp r}
-\left(\frac{\pp\Phi}{\pp r}\right)^2-\frac{\pp^2\Phi}{\pp r^2}\right)b^2
+\mathcal{F}^{-1}\varrho'\nu\left(2\left(\frac{\pp\Phi}{\pp r}+\frac{2}{r}\right)\mathcal{F}
-\frac{6}{r^2}f_R\right)b\lambda \non
&& +\mathcal{F}^{-1}\varrho'\nu e^{2\Lambda}\varrho''\nu b\delta\nu
-\frac{2}{r}\varrho''\nu b\delta\nu
-\frac{\pp\Lambda}{\pp r}\varrho''\nu b\delta\nu
-\varrho''\nu\frac{\pp b}{\pp r}\delta\nu
+\mathcal{F}^{-1}\frac{\pp \mathcal{F}}{\pp r}\varrho''\nu b\delta\nu \non
&& +\mathcal{F}\left(R-\frac{2}{r^2}\right)b\lambda
+\mathcal{F}\frac{b^2}{2f_{RR}}
-\frac{6}{r^2} e^{-2\Lambda}\mathcal{F}^{-1}f_R b\frac{\pp^2b}{\pp r^2} \non
&& -e^{-2\Lambda}
\left(-\frac{2}{r^2}\frac{\pp\Lambda}{\pp r}+\frac{2}{r}\frac{\pp^2\Lambda}{\pp r^2}
+\frac{\pp^2\Lambda}{\pp r^2}\frac{\pp\Phi}{\pp r}+\frac{\pp\Lambda}{\pp r}\frac{\pp^2\Phi}{\pp r^2}
-2\frac{\pp\Phi}{\pp r}\frac{\pp^2\Phi}{\pp r^2}-\frac{\pp^3\Phi}{\pp r^3}\right)b^2 \non
&& -e^{-2\Lambda}\left(\frac{2}{r}\frac{\pp\Lambda}{\pp r}+\frac{\pp\Lambda}{\pp r}\frac{\pp\Phi}{\pp r}-\left(\frac{\pp\Phi}{\pp r}\right)^2
-\frac{2}{r^2}\right)b\frac{\pp b}{\pp r} \non
&& +e^{-2\Lambda}\left(2\left(\frac{\pp^2\Phi}{\pp r^2}-\frac{2}{r^2}\right)\mathcal{F}
+2\left(\frac{\pp\Phi}{\pp r}+\frac{2}{r}\right)\frac{\pp \mathcal{F}}{\pp r}
+\frac{12}{r^3}f_R-\frac{6}{r^2}f_R'\right)b\lambda \non
&& -e^{-2\Lambda}\left(\left(\frac{\pp\Phi}{\pp r}+\frac{2}{r}\right)-\frac{6}{r^2}\mathcal{F}^{-1}f_R\right)
\frac{\pp\mathcal{F}}{\pp r}b\lambda
+e^{-2\Lambda}\left(\left(\frac{\pp\Phi}{\pp r}+\frac{2}{r}\right)-\frac{6}{r^2}\mathcal{F}^{-1}f_R\right)b \non
&& \times \left(-\frac{\pp^2\Phi}{\pp r^2}b-\frac{\pp\Phi}{\pp r}\frac{\pp b}{\pp r}
-2e^{2\Lambda}\frac{\pp\Lambda}{\pp r}\varrho'\nu\xi-e^{2\Lambda}\varrho''\frac{\pp\nu}{\pp r}\nu\xi
-e^{2\Lambda}\varrho'\frac{\pp\nu}{\pp r}\xi-e^{2\Lambda}\varrho'\nu\frac{\pp\xi}{\pp r}\right) \non
&& -e^{-2\Lambda}\left(\frac{\pp\Phi}{\pp r}-2\frac{\pp\Lambda}{\pp r}+\frac{6}{r^2}\mathcal{F}^{-1}f_R\right)
\left(\frac{\pp\Phi}{\pp r}+\frac{2}{r}\right)b\frac{\pp b}{\pp r} \non
&& -e^{-2\Lambda}\left(\frac{\pp\Phi}{\pp r}-2\frac{\pp\Lambda}{\pp r}
+\frac{6}{r^2}\mathcal{F}^{-1}f_R\right)
\left(\frac{2}{r}\frac{\pp\Lambda}{\pp r}
+\frac{\pp\Lambda}{\pp r}\frac{\pp\Phi}{\pp r}-\left(\frac{\pp\Phi}{\pp r}\right)^2
-\frac{\pp^2\Phi}{\pp r^2}\right)b^2 \non
&& \left. +e^{-2\Lambda}\left(\frac{\pp\Phi}{\pp r}-2\frac{\pp\Lambda}{\pp r}
+\frac{6}{r^2}\mathcal{F}^{-1}f_R\right)
\left(2\left(\frac{\pp\Phi}{\pp r}+\frac{2}{r}\right)\mathcal{F}-\frac{6}{r^2}f_R\right)b\lambda
+\mathcal{F}^{-1}\frac{6}{r^2}f_R\varrho''\nu b\delta\nu \right\} \,.
\eean
This is the explicit expression of $\mathcal{P}_{Dyn}=(\psi, \mathcal{T}\psi)$ for $f(R)$ gravity.  Since the integration by parts would be used in the following calculation, we denote the terms associated with $\mathcal{C}_{Dyn}^i$ by $\mathcal{P}_{Dyn}^i$. Now we can read off each $\mathcal{P}_{Dyn}^i$ coefficients from \eq{PDynFinal} one by one:

The first term is the $\left(\frac{\pp b}{\pp r}\right)^2$ term,
\bean
\mathcal{P}_{Dyn}^1 = \int_r (\mathcal{F}^{-2}r^2e^{\Lambda+\Phi}\varrho''\nu^2
+6\mathcal{F}^{-2}e^{-\Lambda+\Phi}f_R)\left(\frac{\pp b}{\pp r}\right)^2 \,,
\eean
hence
\bean \label{PDynC1}
\mathcal{C}_{Dyn}^1 = \mathcal{F}^{-2}r^2e^{\Lambda+\Phi}\varrho''\nu^2+6\mathcal{F}^{-2}e^{-\Lambda+\Phi}f_R \,.
\eean

The second term is the $\left(\frac{\pp\xi}{\pp r}\right)^2$ term,
\bean
\mathcal{P}_{Dyn}^2 = \int_r r^2e^{\Lambda+\Phi}\varrho''\nu^2\left(\frac{\pp\xi}{\pp r}\right)^2 \,,
\eean
so
\bean \label{PDynC2}
\mathcal{C}_{Dyn}^2 = r^2e^{\Lambda+\Phi}\varrho''\nu^2 \,.
\eean

The third term is the $\xi\frac{\pp b}{\pp r}$ term:
\bean
\mathcal{P'}_{Dyn}^3
\eqn \int_r -r^2e^{\Lambda+\Phi}\mathcal{F}^{-1}\varrho'\nu\left(\frac{\pp\Phi}{\pp r}+\frac{2}{r}\right)\xi\frac{\pp b}{\pp r}
+r^2e^{\Lambda+\Phi}\mathcal{F}^{-1}\varrho'\nu\left(2\left(\frac{\pp\Phi}{\pp r}+\frac{2}{r}\right) -\frac{6}{r^2}\mathcal{F}^{-1}f_R\right)\xi\frac{\pp b}{\pp r} \non
&& -r^2e^{\Lambda+\Phi}\mathcal{F}^{-2}\varrho'\nu e^{2\Lambda}\varrho''\nu^2\xi\frac{\pp b}{\pp r}
+2re^{\Lambda+\Phi}\varrho''\nu^2\mathcal{F}^{-1}\xi\frac{\pp b}{\pp r}
+r^2e^{\Lambda+\Phi}\frac{\pp\Lambda}{\pp r}\varrho''\nu^2\mathcal{F}^{-1}\xi\frac{\pp b}{\pp r} \non
&& -r^2e^{\Lambda+\Phi}\varrho'\nu\frac{\pp\Phi}{\pp r}\mathcal{F}^{-1}\xi\frac{\pp b}{\pp r}
+r^2e^{\Lambda+\Phi}\mathcal{F}^{-1}\varrho''\nu^2
\left(\left(\frac{\pp\Lambda}{\pp r}+\frac{2}{r}+\frac{1}{\nu}\frac{\pp\nu}{\pp r}\right)-\mathcal{F}^{-1}e^{2\Lambda}\varrho'\nu\right)
\xi\frac{\pp b}{\pp r} \non
\eqn \int_r \left[r^2e^{\Lambda+\Phi}\mathcal{F}^{-1}\varrho'\nu\left(\left(-\frac{\pp\Phi}{\pp r}+\frac{2}{r}\right) -\frac{6}{r^2}\mathcal{F}^{-1}f_R\right)
-2r^2e^{\Lambda+\Phi}\mathcal{F}^{-1}\left(\frac{\pp\Phi}{\pp r}-\frac{2}{r}-\mathcal{F}^{-1}f_R''\right)\varrho''\nu^2 \right]
\xi\frac{\pp b}{\pp r}  \,.
\eean
And the fourth term is the $b\frac{\pp\xi}{\pp r}$ term:
\bean
\mathcal{P'}_{Dyn}^4
\eqn \int_r -r^2e^{\Lambda+\Phi}\nu\varrho''\frac{\pp\xi}{\pp r}\delta\nu
+r^2e^{\Lambda+\Phi}\mathcal{F}^{-2}\varrho'\nu e^{2\Lambda}\varrho''\nu b\delta\nu
-2re^{\Lambda+\Phi}\mathcal{F}^{-1}\varrho''\nu b\delta\nu
-r^2e^{\Lambda+\Phi}\frac{\pp\Lambda}{\pp r}\mathcal{F}^{-1}\varrho''\nu b\delta\nu \non
&& +r^2\mathcal{F}^{-2}e^{\Lambda+\Phi}\frac{\pp \mathcal{F}}{\pp r}\varrho''\nu b\delta\nu
-r^2\mathcal{F}^{-1}e^{\Lambda+\Phi}\varrho'\nu\left(\left(\frac{\pp\Phi}{\pp r}+\frac{2}{r}\right)-\frac{6}{r^2}\mathcal{F}^{-1}f_R\right)b\frac{\pp\xi}{\pp r}
+r^2\mathcal{F}^{-2}e^{\Lambda+\Phi}\frac{6}{r^2}f_R\varrho''\nu b\delta\nu \non
\eqn \int_r \left[ -2r^2e^{\Lambda+\Phi}\varrho''\nu^2\mathcal{F}^{-1}\frac{\pp\Phi}{\pp r}
-r^2\mathcal{F}^{-1}e^{\Lambda+\Phi}\varrho'\nu\left(\left(\frac{\pp\Phi}{\pp r}+\frac{2}{r}\right) -\frac{6}{r^2}\mathcal{F}^{-1}f_R\right) \right] b\frac{\pp\xi}{\pp r} \,.
\eean
Note that using integration by parts and dropping boundary terms, $\int_r\xi\frac{\pp b}{\pp r}$ terms and $\int_r b\frac{\pp \xi}{\pp r}$ terms can translate to each other. To compare with the coefficients of thermodynamical stability criterion, we can rewrite the coefficients $\mathcal{P'}_{Dyn}^3$ and $\mathcal{P'}_{Dyn}^4$ as
\bean
\mathcal{P'}_{Dyn}^3
\eqn \mathcal{P}_{Dyn}^3+\int_r r^2e^{\Lambda+\Phi}\mathcal{F}^{-1}\varrho'\nu\left(-\frac{\pp\Phi}{\pp r}-\frac{2}{r}
+\frac{6}{r^2}f_R\mathcal{F}^{-1}\right)\xi\frac{\pp b}{\pp r} \non
\eqn \mathcal{P}_{Dyn}^3+\mathcal{P}_{Dyn}^{3\leftrightarrow4}
+\int_r r^2e^{\Lambda+\Phi}\mathcal{F}^{-1}\varrho'\nu\left(\frac{\pp\Phi}{\pp r}+\frac{2}{r}
-\frac{6}{r^2}f_R\mathcal{F}^{-1}\right) b\frac{\pp\xi}{\pp r} \,,
\eean
where
\bean \label{PDynP3}
\mathcal{P}_{Dyn}^3 = \int_r \left[-4e^{\Phi+\Lambda}\mathcal{F}^{-2}\varrho'\nu(f_R-r f_R')
-2r^2e^{\Phi+\Lambda}\mathcal{F}^{-1}\varrho''\nu^2\left(\frac{\pp\Phi}{\pp r}-\frac{2}{r}-f_R''\mathcal{F}^{-1}\right)\right]
\xi\frac{\pp b}{\pp r} \,, \non
\eean
and
\bean \label{PDynP34}
\mathcal{P}_{Dyn}^{3\leftrightarrow4}=\int_r\frac{\pp}{\pp r}\left[r^2e^{\Lambda+\Phi}\mathcal{F}^{-1}\varrho'\nu\left(\frac{\pp\Phi}{\pp r}+\frac{2}{r}
-\frac{6}{r^2}f_R\mathcal{F}^{-1}\right)\right]b \xi \,.
\eean
It is worthy noting that $\mathcal{P}_{Dyn}^{3\leftrightarrow4}$ should be considered when we obtain the coefficient $\mathcal{C}_{Dyn}^8$ of $b\xi$ term in subsequent calculation. We also have
\bean \label{PDynP4}
\mathcal{P}_{Dyn}^{4}
= \mathcal{P'}_{Dyn}^{4}
+\int_r r^2e^{\Lambda+\Phi}\mathcal{F}^{-1}\varrho'\nu\left(\frac{\pp\Phi}{\pp r}+\frac{2}{r}
-\frac{6}{r^2}f_R\mathcal{F}^{-1}\right) b\frac{\pp\xi}{\pp r}
= \int_r -2r^2e^{\Lambda+\Phi}\varrho''\nu^2\mathcal{F}^{-1}\frac{\pp\Phi}{\pp r}b\frac{\pp\xi}{\pp r} \,.
\eean
From \eqs{PDynP3} and \meq{PDynP4} we get
\bean \label{PDynC3}
\mathcal{C}_{Dyn}^3 = -4e^{\Phi+\Lambda}\mathcal{F}^{-2}\varrho'\nu(f_R-r f_R')
-2r^2e^{\Phi+\Lambda}\mathcal{F}^{-1}\varrho''\nu^2\left(\frac{\pp\Phi}{\pp r}-\frac{2}{r}-f_R''\mathcal{F}^{-1}\right)  \,,
\eean
and
\bean \label{PDynC4}
\mathcal{C}_{Dyn}^4 = -2r^2e^{\Lambda+\Phi}\varrho''\nu^2\mathcal{F}^{-1}\frac{\pp\Phi}{\pp r} \,.
\eean

The fifth term is the $\frac{\pp\xi}{\pp r}\frac{\pp b}{\pp r}$ term:
\bean
\mathcal{P}_{Dyn}^5
= \int_r r^2e^{\Lambda+\Phi}\varrho''\nu^2\frac{\pp\xi}{\pp r}\mathcal{F}^{-1}\frac{\pp b}{\pp r}
+r^2e^{\Lambda+\Phi}\mathcal{F}^{-1}\varrho''\nu^2\frac{\pp b}{\pp r}\frac{\pp\xi}{\pp r}
=\int_r 2r^2\mathcal{F}^{-1}e^{\Lambda+\Phi}\varrho''\nu^2\frac{\pp\xi}{\pp r}\frac{\pp b}{\pp r} \,,
\eean
so
\bean \label{PDynC5}
\mathcal{C}_{Dyn}^5 = 2r^2\mathcal{F}^{-1}e^{\Lambda+\Phi}\varrho''\nu^2 \,.
\eean

The calculation of last three terms $\mathcal{P}_{Dyn}^6$, $\mathcal{P}_{Dyn}^7$ and $\mathcal{P}_{Dyn}^8$ are very complicated, so we just show the main steps in our manuscript. Note that $\mathcal{F}=f_R'+\frac{2}{r}f_R$, which yields
\bean
\frac{\pp\mathcal{F}}{\pp r}=f_R''+\frac{2}{r}\mathcal{F}-\frac{6}{r^2}f_R=f_R''+\frac{3}{r}f_R'-\frac{1}{r}\mathcal{F} \,.
\eean
This relation would be used frequently below.

The sixth term is the $b^2$ term, select all terms contain $b^2$ in \eq{PDynFinal}, then direct calculation gives
\bean \label{B24}
\mathcal{P}_{Dyn}^6
\eqn \int_r r^2e^{-\Lambda+\Phi}\mathcal{F}^{-1}\left\{-e^{2\Lambda}\mathcal{F}^{-1}\varrho'\nu
\left(\frac{2}{r}\frac{\pp\Lambda}{\pp r}+\frac{\pp\Lambda}{\pp r}\frac{\pp\Phi}{\pp r}
-\left(\frac{\pp\Phi}{\pp r}\right)^2-\frac{\pp^2\Phi}{\pp r^2}\right)b^2 \right. \non
&& -\frac{\pp\Phi}{\pp r}
\left[ -\frac{2}{r^2}-\frac{6}{r^2}\mathcal{F}^{-1}f_R'+2\left(\frac{\pp\Phi}{\pp r}\right)^2
+\frac{6}{r}\frac{\pp\Phi}{\pp r}-\frac{\pp\Phi}{\pp r}\mathcal{F}^{-1}f_R''+\frac{36}{r^3}\mathcal{F}^{-1}f_R \right. \non
&& \left. -\frac{2}{r}\mathcal{F}^{-1}f_R''-\frac{6}{r^2}\mathcal{F}^{-1}\frac{\pp\Phi}{\pp r}f_R
+\frac{6}{r^2}\mathcal{F}^{-1}f_R\frac{\pp\Lambda}{\pp r}-\frac{72}{r^4}\mathcal{F}^{-2}f_R^2
+\frac{12}{r^2}\mathcal{F}^{-2}f_Rf_R''\right] b^2 \non
&& +e^{2\Lambda}\mathcal{F}\frac{1}{2f_{RR}}b^2
-\left(-\frac{2}{r^2}\frac{\pp\Lambda}{\pp r}+\frac{2}{r}\frac{\pp^2\Lambda}{\pp r^2}
+\frac{\pp^2\Lambda}{\pp r^2}\frac{\pp\Phi}{\pp r}+\frac{\pp\Lambda}{\pp r}\frac{\pp^2\Phi}{\pp r^2}
-2\frac{\pp\Phi}{\pp r}\frac{\pp^2\Phi}{\pp r^2}-\frac{\pp^3\Phi}{\pp r^3}\right)b^2 \non
&& -\left(\left(\frac{\pp\Phi}{\pp r}+\frac{2}{r}\right)-\frac{6}{r^2}\mathcal{F}^{-1}f_R\right)
\frac{\pp^2\Phi}{\pp r^2}b^2 \non
&& -\left(\frac{\pp\Phi}{\pp r}-2\frac{\pp\Lambda}{\pp r}+\frac{6}{r^2}\mathcal{F}^{-1}f_R\right)
\left(\frac{2}{r}\frac{\pp\Lambda}{\pp r}
+\frac{\pp\Lambda}{\pp r}\frac{\pp\Phi}{\pp r}-\left(\frac{\pp\Phi}{\pp r}\right)^2-\frac{\pp^2\Phi}{\pp r^2}\right)b^2 \non
&& +\left[-\frac{2}{r}\frac{\pp\Lambda}{\pp r}-\frac{\pp\Lambda}{\pp r}\frac{\pp\Phi}{\pp r}
-\frac{6}{r^2}\mathcal{F}^{-1}f_R'+3\left(\frac{\pp\Phi}{\pp r}\right)^2
+\frac{6}{r}\frac{\pp\Phi}{\pp r}-\frac{\pp\Phi}{\pp r}\mathcal{F}^{-1}f_R''+\frac{36}{r^3}\mathcal{F}^{-1}f_R \right. \non
&& \left. -\frac{2}{r}\mathcal{F}^{-1}f_R''-\frac{6}{r^2}\mathcal{F}^{-1}\frac{\pp\Phi}{\pp r}f_R
+\frac{6}{r^2}\mathcal{F}^{-1}f_R\frac{\pp\Lambda}{\pp r}
-\frac{72}{r^4}\mathcal{F}^{-2}f_R^2+\frac{12}{r^2}\mathcal{F}^{-2}f_Rf_R''\right] b\frac{\pp b}{\pp r} \non
&& +\left[-\frac{12}{r^2}\mathcal{F}^{-2}\frac{\pp\mathcal{F}}{\pp r}f_R
+\frac{6}{r^2}\mathcal{F}^{-1}\left(-\frac{\pp\Lambda}{\pp r}+\frac{\pp\Phi}{\pp r}\right)f_R
+\frac{6}{r^2}\mathcal{F}^{-1}f_R'\right]b\frac{\pp b}{\pp r} \non
&& -\left(\left(\frac{\pp\Phi}{\pp r}+\frac{2}{r}\right)-\frac{6}{r^2}\mathcal{F}^{-1}f_R\right)
\frac{\pp\Phi}{\pp r}b\frac{\pp b}{\pp r}
-\left(2\frac{\pp\Phi}{\pp r}-\frac{\pp\Lambda}{\pp r}+\frac{6}{r^2}\mathcal{F}^{-1}f_R
-\mathcal{F}^{-1}f_R''\right)
\left(\frac{\pp\Phi}{\pp r}+\frac{2}{r}\right)b\frac{\pp b}{\pp r} \non
&& +\frac{2}{r}e^{2\Lambda}\mathcal{F}^{-1}\frac{\pp\Phi}{\pp r}\varrho''\nu^2b^2
+e^{2\Lambda}\mathcal{F}^{-1}\left(\frac{\pp\Lambda}{\pp r}\frac{\pp\Phi}{\pp r}
+2\left(\frac{\pp\Phi}{\pp r}\right)^2\right)\varrho''\nu^2b^2
-2e^{2\Lambda}\mathcal{F}^{-2}\frac{\pp\mathcal{F}}{\pp r}\frac{\pp\Phi}{\pp r}\varrho''\nu^2b^2 \non
&& \left. +e^{2\Lambda}\mathcal{F}^{-1}\frac{\pp^2\Phi}{\pp r^2}\varrho''\nu^2b^2
-e^{2\Lambda}\mathcal{F}^{-1}\left(\frac{\pp\Phi}{\pp r}\right)^2\frac{\varrho'''\varrho'}{\varrho''}\nu^2b^2
-2e^{2\Lambda}\mathcal{F}^{-1}\left(\frac{\pp\Phi}{\pp r}\right)^2\varrho'\nu b^2 \right\} \,.
\eean
Simplifying \eq{B24} and we obtain the coefficient $\mathcal{C}_{Dyn}^6$ as
\bean
\mathcal{C}_{Dyn}^6 \label{PDynC6}
\eqn -r^2e^{-\Lambda+\Phi}\mathcal{F}^{-1}
\left[10\frac{\pp\Lambda}{\pp r}\frac{\pp\Phi}{\pp r}+3\frac{\pp\Lambda}{\pp r}\left(\frac{\pp\Phi}{\pp r}\right)^2
-3\frac{\pp\Phi}{\pp r}\frac{\pp^2\Phi}{\pp r^2}+\frac{4}{r^2}\frac{\pp\Lambda}{\pp r}
+\frac{2}{r}\frac{\pp^2\Lambda}{\pp r^2} \right. \non
&& \left. +\frac{\pp^2\Lambda}{\pp r^2}\frac{\pp\Phi}{\pp r}
+2\frac{\pp\Lambda}{\pp r}\frac{\pp^2\Phi}{\pp r^2}-\frac{\pp^3\Phi}{\pp r^3}
-\frac{4}{r}\frac{\pp^2\Phi}{\pp r^2}-\frac{2}{r^2}\frac{\pp\Phi}{\pp r}
-\frac{2}{r}\left(\frac{\pp\Lambda}{\pp r}\right)^2-\left(\frac{\pp\Lambda}{\pp r}\right)^2\frac{\pp\Phi}{\pp r} \right] \non
&& +r^2e^{-\Lambda+\Phi}\mathcal{F}^{-2}
\left(\frac{6}{r^2}\frac{\pp\Lambda}{\pp r}+\frac{6}{r}\frac{\pp\Lambda}{\pp r}\frac{\pp\Phi}{\pp r}
-\frac{6}{r}\left(\frac{\pp\Phi}{\pp r}\right)^2-\frac{6}{r}\frac{\pp^2\Phi}{\pp r^2}-\frac{12}{r^2}\frac{\pp\Phi}{\pp r}\right)f_R' \non
&& +r^2e^{-\Lambda+\Phi}\mathcal{F}^{-2}\left(\frac{2}{r}\frac{\pp\Lambda}{\pp r}+\frac{\pp\Lambda}{\pp r}\frac{\pp\Phi}{\pp r}
-\frac{\pp^2\Phi}{\pp r^2}-\frac{4}{r}\frac{\pp\Phi}{\pp r}\right)f_R''+r^2e^{\Lambda+\Phi}\frac{1}{2f_{RR}} \non
&& +r^2e^{\Lambda+\Phi}\mathcal{F}^{-2}\left(\frac{\pp\Lambda}{\pp r}\frac{\pp\Phi}{\pp r}
+2\left(\frac{\pp\Phi}{\pp r}\right)^2+\frac{4}{r}\frac{\pp\Phi}{\pp r}
+\frac{\pp^2\Phi}{\pp r^2}\right)\varrho''\nu^2 \non
&& +2r^2e^{\Lambda+\Phi}\mathcal{F}^{-3}\left(-\frac{3}{r}f_R'-f_R''\right)\frac{\pp\Phi}{\pp r}\varrho''\nu^2
-r^2e^{\Lambda+\Phi}\mathcal{F}^{-2}\left(\frac{\pp\Phi}{\pp r}\right)^2\frac{\varrho'''\varrho'}{\varrho''}\nu^2 \non
&& -2r^2e^{\Lambda+\Phi}\mathcal{F}^{-2}\left(\frac{\pp\Phi}{\pp r}\right)^2\varrho'\nu
+6e^{-\Lambda+\Phi}\mathcal{F}^{-3}\frac{\pp\Phi}{\pp r}\left(3f_R'+r f_R''\right)f_R' \,.
\eean

The seventh term is the $\xi^2$ term, select all terms contain $\xi^2$ in \eq{PDynFinal} we have
\bean
\mathcal{P}_{Dyn}^7
\eqn \int_r -r^2e^{\Lambda+\Phi}\mathcal{F}^{-1}\varrho'\nu\left[2\left(\frac{\pp\Phi}{\pp r}+\frac{2}{r}\right)\mathcal{F} -\frac{6}{r^2}f_R\right]\mathcal{F}^{-1}e^{2\Lambda}\varrho'\nu\xi^2 \non
&& -r^2e^{\Lambda+\Phi}\mathcal{F}^{-1}\varrho'\nu e^{2\Lambda}\varrho''\nu^2\left[\left(\frac{\pp\Lambda}{\pp r}
+\frac{2}{r}+\frac{1}{\nu}\frac{\pp\nu}{\pp r}\right)-\mathcal{F}^{-1}e^{2\Lambda}\varrho'\nu\right]\xi^2 \non
&& +2re^{\Lambda+\Phi}\varrho''\nu^2\left[ \left(\frac{\pp\Lambda}{\pp r}+\frac{2}{r}+\frac{1}{\nu}\frac{\pp\nu}{\pp r}\right)
-\mathcal{F}^{-1}e^{2\Lambda}\varrho'\nu\right]\xi^2 \non
&& +r^2e^{\Lambda+\Phi}\frac{\pp\Lambda}{\pp r}\varrho''\nu^2\left[\left(\frac{\pp\Lambda}{\pp r}+\frac{2}{r}
+\frac{1}{\nu}\frac{\pp\nu}{\pp r}\right)-\mathcal{F}^{-1}e^{2\Lambda}\varrho'\nu\right]\xi^2 \non
&& -r^2e^{\Lambda+\Phi}\varrho'\nu\frac{\pp\Phi}{\pp r}\left[\left(\frac{\pp\Lambda}{\pp r}+\frac{2}{r}+\frac{1}{\nu}\frac{\pp\nu}{\pp r}\right)
-\mathcal{F}^{-1}e^{2\Lambda}\varrho'\nu\right]\xi^2 \non
&& +r^2e^{\Lambda+\Phi}\varrho''\nu^2\left[\left(\frac{\pp\Lambda}{\pp r}+\frac{2}{r}+\frac{1}{\nu}\frac{\pp\nu}{\pp r}\right)
-\mathcal{F}^{-1}e^{2\Lambda}\varrho'\nu\right]\xi\frac{\pp\xi}{\pp r} \non
&& -r^2e^{\Lambda+\Phi}\mathcal{F}^{-1}\varrho'\nu e^{2\Lambda}\varrho''\nu^2\xi\frac{\pp\xi}{\pp r}
+2re^{\Lambda+\Phi}\varrho''\nu^2\xi\frac{\pp\xi}{\pp r} \non
&& +r^2e^{\Lambda+\Phi}\frac{\pp\Lambda}{\pp r}\varrho''\nu^2\xi\frac{\pp\xi}{\pp r}
-r^2e^{\Lambda+\Phi}\varrho'\nu\frac{\pp\Phi}{\pp r}\xi\frac{\pp\xi}{\pp r} \,.
\eean
Using integration by parts we obtain the simplified result of $\mathcal{P}_{Dyn}^7$
\bean
\mathcal{P}_{Dyn}^7
\eqn \int_r r^2e^{\Lambda+\Phi}\varrho'\nu\left[-\frac{\pp\Phi}{\pp r}\frac{\pp\Lambda}{\pp r}-2\left(\frac{\pp\Phi}{\pp r}\right)^2
+2\mathcal{F}^{-1}\frac{\pp\Phi}{\pp r}f_R''-\frac{4}{r}\frac{\pp\Lambda}{\pp r}-\frac{2}{r}\frac{\pp\Phi}{\pp r} \right. \non
&& \left. +\frac{4}{r}\mathcal{F}^{-1}f_R''+\frac{6}{r^2}\mathcal{F}^{-1}f_R\frac{\pp\Lambda}{\pp r}
+\frac{6}{r^2}\mathcal{F}^{-1}f_R\frac{\pp\Phi}{\pp r}
-\frac{6}{r^2}\mathcal{F}^{-2}f_R f_R''+\frac{\pp^2\Phi}{\pp r^2}\right] \non
&& +r^2e^{\Lambda+\Phi}\varrho''\nu^2\left[\frac{\pp\Lambda}{\pp r}\frac{\pp\Phi}{\pp r}+2\left(\frac{\pp\Phi}{\pp r}\right)^2+\frac{\pp^2\Phi}{\pp r^2}
-\frac{2}{r}\frac{\pp\Lambda}{\pp r}-\frac{4}{r}\frac{\pp\Phi}{\pp r}+\frac{2}{r^2} \right. \non
&& \left. -\frac{\pp\Lambda}{\pp r}\mathcal{F}^{-1}f_R''-3\frac{\pp\Phi}{\pp r}\mathcal{F}^{-1}f_R''+2\mathcal{F}^{-2}f_R''^2
-\frac{6}{r^2}\mathcal{F}^{-2}f_R f_R''-\mathcal{F}^{-1}f_R'''+\frac{4}{r}\mathcal{F}^{-1}f_R'' \right] \non
&& +r^2e^{\Lambda+\Phi}\frac{\varrho'''\varrho'}{\varrho''}\nu^2\left[\frac{2}{r}\frac{\pp\Phi}{\pp r}-\left(\frac{\pp\Phi}{\pp r}\right)^2
+\frac{\pp\Phi}{\pp r}f_R''\mathcal{F}^{-1}\right] \,.
\eean
So $\mathcal{C}_{Dyn}^7$ can be obtained. Using \eq{pplusrho} we find that $\mathcal{C}_{Dyn}^7$ can be written as
\bean \label{PDynC7}
\mathcal{C}_{Dyn}^7
\eqn r^2e^{\Lambda+\Phi}\varrho'\nu\left[-\frac{\pp\Phi}{\pp r}\frac{\pp\Lambda}{\pp r}-2\left(\frac{\pp\Phi}{\pp r}\right)^2
+2\mathcal{F}^{-1}\frac{\pp\Phi}{\pp r}f_R''
-\frac{1}{r}\frac{\pp\Lambda}{\pp r}+\frac{1}{r}\frac{\pp\Phi}{\pp r}+\frac{1}{r}\mathcal{F}^{-1}f_R''
-\frac{3}{r}\mathcal{F}^{-2}f_R'\varrho'\nu+\frac{\pp^2\Phi}{\pp r^2}\right] \non
&& +r^2e^{\Lambda+\Phi}\varrho''\nu^2\left[2\frac{\pp\Lambda}{\pp r}\frac{\pp\Phi}{\pp r}
+\left(\frac{\pp\Phi}{\pp r}\right)^2-\frac{6}{r}\frac{\pp\Phi}{\pp r}+\frac{1}{r^2}+\frac{e^{2\Lambda}}{r^2}-e^{2\Lambda}\frac{R}{2}
-\frac{\pp\Lambda}{\pp r}\mathcal{F}^{-1}f_R''-3\frac{\pp\Phi}{\pp r}\mathcal{F}^{-1}f_R''  \right. \non
&& \left. +2\mathcal{F}^{-2}f_R''^2+\frac{2}{r^2}\mathcal{F}^{-2}f_R f_R''-\mathcal{F}^{-1}f_R'''+\frac{4}{r}\mathcal{F}^{-2}f_R''f_R' \right] \non
&& +r^2e^{\Lambda+\Phi}\frac{\varrho'''\varrho'}{\varrho''}\nu^2\left[\frac{2}{r}\frac{\pp\Phi}{\pp r}-\left(\frac{\pp\Phi}{\pp r}\right)^2
+\frac{\pp\Phi}{\pp r}f_R''\mathcal{F}^{-1}\right] \,.
\eean

The last term is the $b\xi$ term, select all $b\xi$ terms in \eq{PDynFinal}, with the addition of \eq{PDynP34}, we have
\bean
\mathcal{P}_{Dyn}^8
\eqn \int_r r^2e^{\Lambda+\Phi}\mathcal{F}^{-1}\left\{-\varrho'\nu
\left(\frac{2}{r}\frac{\pp\Lambda}{\pp r}+\frac{\pp\Lambda}{\pp r}\frac{\pp\Phi}{\pp r}-\left(\frac{\pp\Phi}{\pp r}\right)^2
-\frac{\pp^2\Phi}{\pp r^2}\right) b\xi \right. \non
&& -\varrho'\nu\left(2\left(\frac{\pp\Phi}{\pp r}\right)^2+\frac{4}{r}\frac{\pp\Phi}{\pp r}
-\frac{6}{r^2}\mathcal{F}^{-1}\frac{\pp\Phi}{\pp r}f_R\right)b\xi
+\left(\frac{\pp\Lambda}{\pp r}\frac{\pp\Phi}{\pp r}+\left(\frac{\pp\Phi}{\pp r}\right)^2
-\mathcal{F}^{-1}\frac{\pp\Phi}{\pp r}f_R''\right)\varrho''\nu^2b\xi \non
&& -\frac{2}{r}\frac{\pp\Phi}{\pp r}\varrho''\nu^2 b\xi
-\frac{\pp\Lambda}{\pp r}\frac{\pp\Phi}{\pp r}\varrho''\nu^2 b\xi
+\varrho'\nu\left(\frac{\pp\Phi}{\pp r}\right)^2b\xi \non
&& -\left(-\frac{2}{r^2}+\frac{4}{r}\frac{\pp\Lambda}{\pp r}-\frac{4}{r}\frac{\pp\Phi}{\pp r}
+2\frac{\pp\Lambda}{\pp r}\frac{\pp\Phi}{\pp r}-2\left(\frac{\pp\Phi}{\pp r}\right)^2-2\frac{\pp^2\Phi}{\pp r^2}\right)\varrho'\nu b\xi \non
&& -\mathcal{F}^{-1}\left[2\left(\frac{\pp^2\Phi}{\pp r^2}-\frac{2}{r^2}\right)\mathcal{F}
+\left(\frac{\pp\Phi}{\pp r}+\frac{2}{r}\right)\frac{\pp \mathcal{F}}{\pp r}
+\frac{12}{r^3}f_R-\frac{6}{r^2}f_R'+\frac{6}{r^2}\mathcal{F}^{-1}\frac{\pp\mathcal{F}}{\pp r}f_R\right]\varrho'\nu b\xi \non
&& +\left(\left(\frac{\pp\Phi}{\pp r}+\frac{2}{r}\right)-\frac{6}{r^2}\mathcal{F}^{-1}f_R\right)
\left(-2\frac{\pp\Lambda}{\pp r}\varrho'\nu+\frac{\pp\Phi}{\pp r}\varrho'\nu
+\frac{\varrho'^2}{\varrho''}\frac{\pp\Phi}{\pp r}\right)b\xi \non
&& -\mathcal{F}^{-1}\left[2\frac{\pp\Phi}{\pp r}-\frac{\pp\Lambda}{\pp r}+\frac{6}{r^2}\mathcal{F}^{-1}f_R
-\mathcal{F}^{-1}f_R''\right]
\left[2\left(\frac{\pp\Phi}{\pp r}+\frac{2}{r}\right)\mathcal{F}-\frac{6}{r^2}f_R\right]\varrho'\nu b\xi \non
&& -\mathcal{F}^{-1}\frac{6}{r^2}f_R\varrho''\nu^2
\left(\frac{2}{r}+\frac{\varrho'}{\varrho''\nu}\frac{\pp\Phi}{\pp r}-\frac{\pp\Phi}{\pp r}+\mathcal{F}^{-1}f_R''\right)b\xi
-\mathcal{F}^{-1}\varrho'\nu e^{2\Lambda}\varrho''\nu^2
\left(\frac{2}{r}+\frac{\varrho'}{\varrho''\nu}\frac{\pp\Phi}{\pp r}-\frac{\pp\Phi}{\pp r}+\mathcal{F}^{-1}f_R''\right)b\xi \non
&& +\frac{2}{r}\varrho''\nu^2
\left(\frac{2}{r}+\frac{\varrho'}{\varrho''\nu}\frac{\pp\Phi}{\pp r}-\frac{\pp\Phi}{\pp r}+\mathcal{F}^{-1}f_R''\right)b\xi
+\frac{\pp\Lambda}{\pp r}\varrho''\nu^2
\left(\frac{2}{r}+\frac{\varrho'}{\varrho''\nu}\frac{\pp\Phi}{\pp r}-\frac{\pp\Phi}{\pp r}+\mathcal{F}^{-1}f_R''\right)b\xi \non
&& \left. -\mathcal{F}^{-1}\frac{\pp \mathcal{F}}{\pp r}\varrho''\nu^2
\left(\frac{2}{r}+\frac{\varrho'}{\varrho''\nu}\frac{\pp\Phi}{\pp r}-\frac{\pp\Phi}{\pp r}+\mathcal{F}^{-1}f_R''\right)b\xi \right\}
+\mathcal{P}_{Dyn}^{3\leftrightarrow4}  \,.
\eean
After some calculations we obtain
\bean
\mathcal{P}_{Dyn}^8
\eqn \int_r 2r^2e^{\Lambda+\Phi}\mathcal{F}^{-1}\frac{\pp\Phi}{\pp r}
\left(\frac{\pp\Phi}{\pp r}-\frac{2}{r}-\mathcal{F}^{-1}f_R''\right)\varrho''\nu^2b\xi \non
&& -r^2e^{\Lambda+\Phi}\left(\frac{4}{r}\frac{\pp\Lambda}{\pp r}+2\frac{\pp\Lambda}{\pp r}\frac{\pp\Phi}{\pp r}-2\frac{\pp^2\Phi}{\pp r^2}
+\frac{2}{r}\frac{\pp\Phi}{\pp r}\right)\mathcal{F}^{-1}\varrho'\nu b\xi
-r^2e^{\Lambda+\Phi}\frac{6}{r}\mathcal{F}^{-2}\frac{\pp\Phi}{\pp r}f_R'\varrho'\nu b\xi \,,
\eean
which yields
\bean
\mathcal{C}_{Dyn}^8 \label{PDynC8}
\eqn 2r^2e^{\Lambda+\Phi}\mathcal{F}^{-1}\frac{\pp\Phi}{\pp r}
\left(\frac{\pp\Phi}{\pp r}-\frac{2}{r}-\mathcal{F}^{-1}f_R''\right)\varrho''\nu^2 \non
&& -r^2e^{\Lambda+\Phi}\left(\frac{4}{r}\frac{\pp\Lambda}{\pp r}+2\frac{\pp\Lambda}{\pp r}\frac{\pp\Phi}{\pp r}-2\frac{\pp^2\Phi}{\pp r^2}
+\frac{2}{r}\frac{\pp\Phi}{\pp r}\right)\mathcal{F}^{-1}\varrho'\nu
-r^2e^{\Lambda+\Phi}\frac{6}{r}\mathcal{F}^{-2}\frac{\pp\Phi}{\pp r}f_R'\varrho'\nu \,.
\eean

Substituting these coefficient $\mathcal{C}_{Dyn}^i$ into Eq. (\ref{DynResult}) gives the dynamical stability criterion for perfect fluid in $f(R)$ theories.

Degenerate to general relativity, $f(R)=R$, hence $b=\delta f_R=0$, and only $\mathcal{C}_{Dyn}^1\hsp_{degenerate}$ and $\mathcal{C}_{Dyn}^7\hsp_{degenerate}$ remain. It is easy to check that
\bean
\mathcal{P}_{Dyn}\hsp_{degenerate}=\int_r \mathcal{C}_{Dyn}^1\hsp_{degenerate}\left(\frac{\pp\xi}{\pp r}\right)^2
+\mathcal{C}_{Dyn}^7\hsp_{degenerate}\xi^2
\eean
is the dynamical stability criterion given by Eq.(97) of Ref. \cite{wald2007}.

\section{Detailed calculation and result of Thermodynamical stability criterion}
In this appendix, we will show the detailed calculations of the explicitly form of thermodynamical stability criterion, $\delta^2S$. From \eqs{delsgt} and \meq{firstvariationrho}, we obtain the second variation of $\sgt$ as
\bean
\delta^2\sgt=r^2e^{\Lambda}\lambda^2+r^2e^{\Lambda}\delta\lambda \,,
\eean
and the second variation of $\rho$ as
\bean \label{secondvariationrho}
\delta^2\rho \eqn \frac{\delta b}{r^2}+\frac{4e^{-2\Lambda}}{r^2}\left(2r\frac{\pp\Lambda}{\pp r}-1\right)f_R\lambda^2
-\frac{2e^{-2\Lambda}}{r}f_R\frac{\pp}{\pp r}\lambda^2
-\frac{4e^{-2\Lambda}}{r^2}\left(2r\frac{\pp\Lambda}{\pp r}-1\right)b\lambda
-\frac{2e^{-2\Lambda}}{r^2}\left(2r\frac{\pp\Lambda}{\pp r}-1\right)f_R\delta\lambda \non
&& +\frac{4e^{-2\Lambda}}{r}b\frac{\pp\lambda}{\pp r}+\frac{e^{-2\Lambda}}{r^2}\left(2r\frac{\pp\Lambda}{\pp r}-1\right)\delta b
-\frac{2e^{-2\Lambda}}{r}f_R\frac{\pp}{\pp r}\lambda^2+\frac{2e^{-2\Lambda}}{r}f_R\frac{\pp}{\pp r}\delta\lambda
-\frac{b}{2}\delta R-\frac{R}{2}\delta b \non
&& -4e^{-2\Lambda}f_R''\lambda^2
+4e^{-2\Lambda}b''\lambda+2e^{-2\Lambda}f_R''\delta\lambda
-e^{-2\Lambda}\delta b''+4e^{-2\Lambda}\left(\frac{\pp\Lambda}{\pp r}-\frac{2}{r}\right)f_R'\lambda^2
-2e^{-2\Lambda}f_R'\frac{\pp}{\pp r}\lambda^2 \non
&& -2e^{-2\Lambda}\left(\frac{\pp\Lambda}{\pp r}-\frac{2}{r}\right)b'\lambda
-2e^{-2\Lambda}\left(\frac{\pp\Lambda}{\pp r}-\frac{2}{r}\right)f_R'\delta\lambda
+e^{-2\Lambda}f_R'\frac{\pp}{\pp r}\delta\lambda
+2e^{-2\Lambda}\left(\frac{\pp\lambda}{\pp r}\right)\cdot b' \non
&& -2e^{-2\Lambda}\left(\frac{\pp\Lambda}{\pp r}-\frac{2}{r}\right)b'\lambda
+e^{-2\Lambda}\left(\frac{\pp\Lambda}{\pp r}-\frac{2}{r}\right)\delta b' \,.
\eean
Now we can calculate the terms in the righthand side of \eq{delta2S} one by one. Note that in spherical symmetry case $\int_C$ becomes $\int_r$. The first term in the righthand side of \eq{delta2S} can be calculated as
\bean
\int_r\frac{2}{T}\delta\rho\delta\sgt
\eqn \int_r 2e^{\Phi+\Lambda}r^2 \left[\frac{1}{r^2}b\lambda
-\frac{2e^{-2\Lambda}}{r^2}\left(2r\frac{\pp\Lambda}{\pp r}-1\right)f_R\lambda^2
+\frac{e^{-2\Lambda}}{r}f_R\frac{\pp}{\pp r}\lambda^2 \right. \non
&& +\frac{e^{-2\Lambda}}{r^2}\left(2r\frac{\pp\Lambda}{\pp r}-1\right)b\lambda
-\frac{R}{2}b\lambda+2e^{-2\Lambda}f_R''\lambda^2-e^{-2\Lambda}b''\lambda \non
&& \left. -2e^{-2\Lambda}\left(\frac{\pp\Lambda}{\pp r}-\frac{2}{r}\right)f_R'\lambda^2
+\frac{1}{2}e^{-2\Lambda}f_R'\frac{\pp}{\pp r}\lambda^2
+e^{-2\Lambda}\left(\frac{\pp\Lambda}{\pp r}-\frac{2}{r}\right)b'\lambda \right] \,. \non
\eean
Using integration by parts we obtain
\bean \label{Thermofirst}
\int_r \frac{2}{T}\delta\rho\delta\sgt
\eqn \int_r e^{\Phi+\Lambda}r^2
\left[e^{-2\Lambda}\left(\frac{4}{r}\frac{\pp\Phi}{\pp r}-2\frac{\pp\Lambda}{\pp r}\frac{\pp\Phi}{\pp r}
+2\left(\frac{\pp\Phi}{\pp r}\right)^2+2\frac{\pp^2\Phi}{\pp r^2}\right)b\lambda  \right. \non
&& +2e^{-2\Lambda}\frac{\pp\Phi}{\pp r}b'\lambda+2e^{-2\Lambda}b'\frac{\pp\lambda}{\pp r}  \non
&& \left. +\left(-3(p+\rho)+e^{-2\Lambda}\left(\frac{4}{r}\frac{\pp\Phi}{\pp r}+\frac{2}{r^2}\right)f_R
+e^{-2\Lambda}\left(2\frac{\pp\Phi}{\pp r}+\frac{4}{r}\right)f_R'\right)\lambda^2  \right] \,. \non
\eean

With \eq{secondvariationrho}, the second term in the righthand side of \eq{delta2S} becomes
\bean
\int_r\frac{1}{T}\sgt\delta^2\rho \eqn
\int_r e^{\Phi+\Lambda}r^2 \cdot \left[\frac{\delta b}{r^2}+\frac{4e^{-2\Lambda}}{r^2}\left(2r\frac{\pp\Lambda}{\pp r}-1\right)f_R\lambda^2
-\frac{2e^{-2\Lambda}}{r}f_R\frac{\pp}{\pp r}\lambda^2-\frac{4e^{-2\Lambda}}{r^2}\left(2r\frac{\pp\Lambda}{\pp r}-1\right)b\lambda \right. \non
&& -\frac{2e^{-2\Lambda}}{r^2}\left(2r\frac{\pp\Lambda}{\pp r}-1\right)f_R\delta\lambda
+\frac{4e^{-2\Lambda}}{r}b\frac{\pp\lambda}{\pp r}+\frac{e^{-2\Lambda}}{r^2}\left(2r\frac{\pp\Lambda}{\pp r}-1\right)\delta b
-\frac{2e^{-2\Lambda}}{r}f_R\frac{\pp}{\pp r}\lambda^2 \non
&& +\frac{2e^{-2\Lambda}}{r}f_R\frac{\pp}{\pp r}\delta\lambda
-\frac{b}{2}\delta R-\frac{R}{2}\delta b-4e^{-2\Lambda}f_R''\lambda^2
+4e^{-2\Lambda}b''\lambda+2e^{-2\Lambda}f_R''\delta\lambda-e^{-2\Lambda}\delta b'' \non
&& +4e^{-2\Lambda}\left(\frac{\pp\Lambda}{\pp r}-\frac{2}{r}\right)f_R'\lambda^2
-2e^{-2\Lambda}f_R'\frac{\pp}{\pp r}\lambda^2
-2e^{-2\Lambda}\left(\frac{\pp\Lambda}{\pp r}-\frac{2}{r}\right)b'\lambda
-2e^{-2\Lambda}\left(\frac{\pp\Lambda}{\pp r}-\frac{2}{r}\right)f_R'\delta\lambda \non
&& \left. +e^{-2\Lambda}f_R'\frac{\pp}{\pp r}\delta\lambda
+2e^{-2\Lambda}\left(\frac{\pp \lambda}{\pp r}\right)\cdotp b'
-2e^{-2\Lambda}\left(\frac{\pp\Lambda}{\pp r}-\frac{2}{r}\right)b'\lambda
+e^{-2\Lambda}\left(\frac{\pp\Lambda}{\pp r}-\frac{2}{r}\right)\delta b' \right] \,.
\eean
Using integration by parts, we obtain the simplified expression as
\bean \label{Thermosecond}
\int_r\frac{1}{T}\sgt\delta^2\rho \eqn
\int_r e^{\Phi+\Lambda}r^2 \cdot \left\{ \left[\frac{4e^{-2\Lambda}}{r}\left(\frac{\pp\Lambda}{\pp r}+\frac{\pp\Phi}{\pp r}\right)f_R
+2e^{-2\Lambda}\left(\frac{\pp\Lambda}{\pp\Phi}+\frac{\pp\Phi}{\pp r}\right)f_R'-2e^{-2\Lambda}f_R''\right]\lambda^2 \right. \non
&& +\left[e^{-2\Lambda}f_R''-e^{-2\Lambda}f_R'\left(\frac{\pp\Phi}{\pp r}+\frac{\pp\Lambda}{\pp r}\right)
-2e^{-2\Lambda}\left(\frac{\pp\Lambda}{\pp r}+\frac{\pp\Phi}{\pp r}\right)\frac{f_R}{r}\right]\delta\lambda \non
&& \left. -\frac{4e^{-2\Lambda}}{r^2}\left(2r\frac{\pp\Lambda}{\pp r}-1\right)b\lambda
+\frac{4e^{-2\Lambda}}{r}b\frac{\pp\lambda}{\pp r}-\frac{b}{2}\delta R-2e^{-2\Lambda}b'\frac{\pp\lambda}{\pp r}
-4e^{-2\Lambda}\frac{\pp\Phi}{\pp r}b'\lambda \right\} \,.
\eean

The third term in the righthand side of \eq{delta2S} is $\int_r\frac{1}{T}(p+\rho)\delta^2\sgt$, which can be written as
\bean \label{Thermothird}
\int_r\frac{1}{T}(p+\rho)\delta^2\sgt =
\int_re^{\Phi}(p+\rho)[e^{\Lambda}r^2\lambda^2+e^{\Lambda}r^2\delta\lambda] \,.
\eean
And the fourth term in the righthand side of \eq{delta2S} is
\bean \label{Thermofourth}
\int_r-\frac{1}{T}\frac{\delta p\delta\rho}{p+\rho}\sgt = -\int_r e^{\Phi+\Lambda}r^2\varrho''(\delta\nu)^2 \,.
\eean

Together with \eqs{Thermofirst}, \meq{Thermosecond}, \meq{Thermothird} and \meq{Thermofourth}, the second variation of total entropy takes the form
\bean \label{delta2SFinal}
\delta^2S \eqn \int_r \left(-4e^{\Phi-\Lambda}r\frac{\pp\Lambda}{\pp r}-2e^{\Phi-\Lambda}r^2\frac{\pp\Lambda}{\pp r}\frac{\pp\Phi}{\pp r}
+2e^{\Phi-\Lambda}r^2\left(\frac{\pp\Phi}{\pp r}\right)^2+2e^{\Phi-\Lambda}r^2\frac{\pp^2\Phi}{\pp r^2}\right)
\mathcal{F}^{-1}\left[-\varrho'\nu e^{2\Lambda}\xi b+\frac{1}{2}\frac{\pp}{\pp r}b^2-\frac{\pp\Phi}{\pp r}b^2 \right] \non
&& +\left[\left(4e^{\Phi-\Lambda}r\frac{\pp\Phi}{\pp r}+2e^{\Phi-\Lambda}\right)f_R
+\left(2e^{\Phi-\Lambda}r^2\frac{\pp\Phi}{\pp r}+4e^{\Phi-\Lambda}r\right)f_R'\right] \non
&& \times \mathcal{F}^{-2}\left[(\varrho'\nu)^2e^{4\Lambda}\xi^2
+\left(\frac{\pp}{\pp r}b\right)^2+\left(\frac{\pp\Phi}{\pp r}\right)^2b^2
-2\varrho'\nu e^{2\Lambda}\xi\frac{\pp}{\pp r}b+2\varrho'\nu e^{2\Lambda}\xi\frac{\pp\Phi}{\pp r}b
-2\frac{\pp\Phi}{\pp r}b\frac{\pp }{\pp r}b \right]
-e^{\Phi+\Lambda}r^2\frac{b^2}{2f_{RR}} \non
&& -\left( 2e^{\Phi-\Lambda}\frac{\pp\Phi}{\pp r}r^2+4e^{\Phi-\Lambda}r \right)
\times \mathcal{F}^{-1}\left[-\varrho'\nu e^{2\Lambda}\xi\frac{\pp b}{\pp r}
+\left(\frac{\pp b}{\pp r}\right)^2-\frac{1}{2}\frac{\pp\Phi}{\pp r}\frac{\pp}{\pp r}b^2 \right] \non
&& -e^{\Phi+\Lambda}r^2 \varrho''
\left[-\nu\frac{\pp\xi}{\pp r}-\mathcal{F}^{-1}\nu\frac{\pp b}{\pp r}+\mathcal{F}^{-1}\nu\frac{\pp\Phi}{\pp r}b
+\left(\nu\frac{\pp\Phi}{\pp r}
+\frac{\varrho'}{\varrho''}\frac{\pp\Phi}{\pp r}-\frac{2}{r}\nu-\nu f_R''\mathcal{F}^{-1}\right)\xi
 \right]^2 \,.
\eean
The $\delta\nu$ terms in \eq{Thermofourth} can be calculated by \eqs{delnu} and \meq{lambda}. It is worthy noting that all second variation of variables, such as $\delta b$ and $\delta\lambda$, vanish. That is because we assume that the system state is deviated only slightly from equilibrium state. Now the coefficients $\mathcal{C}_{Thermo}^1$ to $\mathcal{C}_{Thermo}^8$ can be directly read off from \eq{delta2SFinal}.

The first term is the $\left(\frac{\pp b}{\pp r}\right)^2$ term
\bean \label{STherC1}
\mathcal{C}_{Thermo}^1 \eqn \left[\left(4e^{\Phi-\Lambda}r\frac{\pp\Phi}{\pp r}+2e^{\Phi-\Lambda}\right)f_R
+\left(2e^{\Phi-\Lambda}r^2\frac{\pp\Phi}{\pp r}+4e^{\Phi-\Lambda}r\right)f_R'\right]\mathcal{F}^{-2} \non
&& -\left( 2e^{\Phi-\Lambda}\frac{\pp\Phi}{\pp r}r^2+4e^{\Phi-\Lambda}r \right)\mathcal{F}^{-1}
-e^{\Phi+\Lambda}r^2\varrho''(\mathcal{F}^{-1}\nu)^2 \non
\eqn -\mathcal{F}^{-2}(6e^{\Phi-\Lambda}f_R+e^{\Phi+\Lambda}\varrho'' r^2\nu^2) \,.
\eean

The second term is the $\left(\frac{\pp \xi}{\pp r}\right)^2$ term
\bean \label{STherC2}
\mathcal{C}_{Thermo}^2 = -e^{\Phi+\Lambda}r^2\varrho''\nu^2 \,.
\eean

The third term is the $\xi\frac{\pp b}{\pp r}$ term
\bean \label{STherC3}
\mathcal{C}_{Thermo}^3 \eqn 2e^{\Phi-\Lambda}\left[\left(2r\frac{\pp\Phi}{\pp r}+1\right)f_R
+\left(r^2\frac{\pp\Phi}{\pp r}+2r\right)f_R'\right]
\mathcal{F}^{-2}\left(-2\varrho'\nu e^{2\Lambda}\right) \non
&& -\left(2e^{\Phi-\Lambda}\frac{\pp\Phi}{\pp r}r^2+4e^{\Phi-\Lambda}r\right)
\mathcal{F}^{-2}\left(\frac{2}{r}f_R+f_R'\right)(-\varrho'\nu e^{2\Lambda}) \non
&& -2e^{\Phi+\Lambda}r^2\varrho''\left(\nu\frac{\pp\Phi}{\pp r}
+\frac{\varrho'}{\varrho''}\frac{\pp\Phi}{\pp r}
-\frac{2}{r}\nu-\nu f_R''\mathcal{F}^{-1}\right)(-\mathcal{F}^{-1}\nu) \non
\eqn 4e^{\Phi+\Lambda}\mathcal{F}^{-2}\varrho'\nu(f_R-r f_R')+2r^2e^{\Phi+\Lambda}\mathcal{F}^{-1}\varrho''\nu^2\left(\frac{\pp\Phi}{\pp r}-\frac{2}{r}-f_R''\mathcal{F}^{-1}\right) \,.
\eean

The fourth term is the $b\frac{\pp\xi}{\pp r}$ term
\bean \label{STherC4}
\mathcal{C}_{Thermo}^4 = 2e^{\Phi+\Lambda}r^2\mathcal{F}^{-1}\varrho''\nu^2\frac{\pp\Phi}{\pp r} \,.
\eean

The fifth term is the $\frac{\pp\xi}{\pp r}\frac{\pp b}{\pp r}$ term
\bean \label{STherC5}
\mathcal{C}_{Thermo}^5 \eqn -2e^{\Phi+\Lambda}r^2\mathcal{F}^{-1}\varrho''\nu^2 \,.
\eean

Since integration by parts will be used when we calculate the sixth term $b^2$ and the seventh term $\xi^2$. Similarly to appendix (B),
we denote the terms associated with $\mathcal{C}_{Thermo}^6$ and $\mathcal{C}_{Thermo}^7$ by $\mathcal{P}_{Thermo}^6$ and $\mathcal{P}_{Thermo}^7$, respectively. Select all terms contain $b^2$ in \eq{delta2SFinal}, we get the sixth term as
\bean \label{C15}
\mathcal{P}_{Thermo}^6
\eqn \int_r e^{\Phi-\Lambda}\left(-2r\frac{\pp\Lambda}{\pp r}
-r^2\frac{\pp\Lambda}{\pp r}\frac{\pp\Phi}{\pp r}
+r^2\left(\frac{\pp\Phi}{\pp r}\right)^2
+r^2\frac{\pp^2\Phi}{\pp r^2}\right) \mathcal{F}^{-1}\frac{\pp}{\pp r}b^2 \non
&& +2e^{\Phi-\Lambda}\left(-2r\frac{\pp\Lambda}{\pp r}
-r^2\frac{\pp\Lambda}{\pp r}\frac{\pp\Phi}{\pp r}
+r^2\left(\frac{\pp\Phi}{\pp r}\right)^2
+r^2\frac{\pp^2\Phi}{\pp r^2}\right)
\mathcal{F}^{-1}\left(-\frac{\pp\Phi}{\pp r}\right)b^2 \non
&& +2e^{\Phi-\Lambda}\left[\left(2r\frac{\pp\Phi}{\pp r}+1\right)f_R
+\left(r^2\frac{\pp\Phi}{\pp r}+2r\right)f_R'\right]\mathcal{F}^{-2}
\left(\frac{\pp\Phi}{\pp r}\right)^2b^2 \non
&& -2e^{\Phi-\Lambda}\left[\left(2r\frac{\pp\Phi}{\pp r}+1\right)f_R
+\left(r^2\frac{\pp\Phi}{\pp r}+2r\right)f_R'\right]
\mathcal{F}^{-2}\frac{\pp\Phi}{\pp r}\frac{\pp}{\pp r}b^2 \non
&& -e^{\Phi+\Lambda}r^2\frac{b^2}{2f_{RR}}
-\left(e^{\Phi-\Lambda}\frac{\pp\Phi}{\pp r}r^2+2e^{\Phi-\Lambda}r\right)
\mathcal{F}^{-1}\left(-\frac{\pp\Phi}{\pp r}\right)\frac{\pp}{\pp r}b^2 \non
&& -e^{\Phi+\Lambda}r^2\varrho''\left(\mathcal{F}^{-1}\nu\frac{\pp\Phi}{\pp r}\right)^2 b^2
-2e^{\Phi+\Lambda}r^2\varrho''\left(-\mathcal{F}^{-1}\nu\frac{\pp b}{\pp r}\right)
\left(\mathcal{F}^{-1}\nu\frac{\pp\Phi}{\pp r}b\right) \,.
\eean
Simplifying \eq{C15} yields
\bean
\mathcal{P}_{Thermo}^6
\eqn \int_r e^{\Phi-\Lambda}\mathcal{F}^{-1}\left[4r\frac{\pp\Lambda}{\pp r}+2r^2\frac{\pp\Lambda}{\pp r}\frac{\pp\Phi}{\pp r}
-2r^2\frac{\pp^2\Phi}{\pp r^2}+r\frac{\pp\Phi}{\pp r}\right]\frac{\pp\Phi}{\pp r}b^2 \non
&& -e^{\Phi+\Lambda}r^2\frac{b^2}{2f_{RR}}
-e^{\Phi+\Lambda}r^2\varrho''\mathcal{F}^{-2}\nu^2\left(\frac{\pp\Phi}{\pp r}\right)^2b^2
+3e^{\Phi-\Lambda}r\mathcal{F}^{-2}\left(\frac{\pp\Phi}{\pp r}\right)^2f_R'b^2 \non
&& -e^{\Phi-\Lambda}\left[2r\frac{\pp\Lambda}{\pp r}+r^2\frac{\pp\Lambda}{\pp r}\frac{\pp\Phi}{\pp r}-r^2\frac{\pp^2\Phi}{\pp r^2}
-r\frac{\pp\Phi}{\pp r}\right]\mathcal{F}^{-1}\frac{\pp}{\pp r}b^2 \non
&& -3rf_R'e^{\Phi-\Lambda}\mathcal{F}^{-2}\frac{\pp\Phi}{\pp r}\frac{\pp}{\pp r}b^2 +e^{\Phi+\Lambda}r^2\varrho''\mathcal{F}^{-2}\nu^2\frac{\pp\Phi}{\pp r}\frac{\pp}{\pp r}b^2 \,.
\eean
Using integration by parts, and after many calculations, we obtain the simplified result takes the form
\bean
\mathcal{P}_{Thermo}^6 \eqn \int_re^{\Phi-\Lambda}\left[10r\frac{\pp\Lambda}{\pp r}\frac{\pp\Phi}{\pp r}
+3r^2\frac{\pp\Lambda}{\pp r}\left(\frac{\pp\Phi}{\pp r}\right)^2
-3r^2\frac{\pp^2\Phi}{\pp r^2}\frac{\pp\Phi}{\pp r}+4\frac{\pp\Lambda}{\pp r}+2r\frac{\pp^2\Lambda}{\pp r^2}
+r^2\frac{\pp^2\Lambda}{\pp r^2}\frac{\pp\Phi}{\pp r} \right. \non
&& \left.+2r^2\frac{\pp\Lambda}{\pp r}\frac{\pp^2\Phi}{\pp r^2}-r^2\frac{\pp^3\Phi}{\pp r^3}
-2\frac{\pp\Phi}{\pp r}-4r\frac{\pp^2\Phi}{\pp r^2}-2r\left(\frac{\pp\Lambda}{\pp r}\right)^2
-r^2\left(\frac{\pp\Lambda}{\pp r}\right)^2\frac{\pp\Phi}{\pp r}\right]\mathcal{F}^{-1}b^2 \non
&& +e^{\Phi-\Lambda}\left(-2r\frac{\pp\Lambda}{\pp r}-r^2\frac{\pp\Lambda}{\pp r}\frac{\pp\Phi}{\pp r}
+r^2\frac{\pp^2\Phi}{\pp r^2}+4r\frac{\pp\Phi}{\pp r}\right)\mathcal{F}^{-2}f_R''b^2 \non
&& +e^{\Phi-\Lambda}\left(-6\frac{\pp\Lambda}{\pp r}-3r\frac{\pp\Lambda}{\pp r}\frac{\pp\Phi}{\pp r}
+3r\frac{\pp^2\Phi}{\pp r^2}+12\frac{\pp\Phi}{\pp r}\right)\mathcal{F}^{-2}f_R'b^2 \non
&& +f_R'e^{\Phi-\Lambda}\left(6r\left(\frac{\pp\Phi}{\pp r}\right)^2
-3r\frac{\pp\Phi}{\pp r}\frac{\pp\Lambda}{\pp r}\right)\mathcal{F}^{-2}b^2
+3rf_R'e^{\Phi-\Lambda}\mathcal{F}^{-2}\frac{\pp^2\Phi}{\pp r^2}b^2 \non
&& -e^{\Phi+\Lambda}r^2\varrho''\mathcal{F}^{-2}\nu^2\left[\frac{\pp\Lambda}{\pp r}\frac{\pp\Phi}{\pp r}
+2\left(\frac{\pp\Phi}{\pp r}\right)^2+\frac{4}{r}\frac{\pp\Phi}{\pp r}+\frac{\pp^2\Phi}{\pp r^2}\right]b^2 \non
&& -6rf_R'e^{\Phi-\Lambda}\mathcal{F}^{-3}\left(\frac{3}{r}f_R'+f_R''\right)\frac{\pp\Phi}{\pp r}b^2 +e^{\Phi+\Lambda}r^2\frac{\varrho'''\varrho'}{\varrho''}\mathcal{F}^{-2}\nu^2\left(\frac{\pp\Phi}{\pp r}\right)^2b^2 \non
&& +2e^{\Phi+\Lambda}r^2\mathcal{F}^{-3}\left(\frac{3}{r}f_R'+f_R''\right)
\varrho''\nu^2\frac{\pp\Phi}{\pp r}b^2
+2e^{\Phi+\Lambda}r^2\mathcal{F}^{-2}\varrho'\nu\left(\frac{\pp\Phi}{\pp r}\right)^2b^2
-e^{\Phi+\Lambda}r^2\frac{b^2}{2f_{RR}} \,.
\eean
which means that $\mathcal{C}_{Thermo}^6$ can be written as
\bean \label{STherC6}
\mathcal{C}_{Thermo}^6 \eqn r^2e^{\Phi-\Lambda}\mathcal{F}^{-1}\left[\frac{10}{r}\frac{\pp\Lambda}{\pp r}\frac{\pp\Phi}{\pp r}
+3\frac{\pp\Lambda}{\pp r}\left(\frac{\pp\Phi}{\pp r}\right)^2
-3\frac{\pp^2\Phi}{\pp r^2}\frac{\pp\Phi}{\pp r}+\frac{4}{r^2}\frac{\pp\Lambda}{\pp r}
+\frac{2}{r}\frac{\pp^2\Lambda}{\pp r^2} \right. \non
&& \left. +\frac{\pp^2\Lambda}{\pp r^2}\frac{\pp\Phi}{\pp r}+2\frac{\pp\Lambda}{\pp r}\frac{\pp^2\Phi}{\pp r^2}-\frac{\pp^3\Phi}{\pp r^3}
-\frac{2}{r^2}\frac{\pp\Phi}{\pp r}-\frac{4}{r}\frac{\pp^2\Phi}{\pp r^2}-\frac{2}{r}\left(\frac{\pp\Lambda}{\pp r}\right)^2
-\left(\frac{\pp\Lambda}{\pp r}\right)^2\frac{\pp\Phi}{\pp r}\right] \non
&& +e^{\Phi-\Lambda}\mathcal{F}^{-2}\left(-2r\frac{\pp\Lambda}{\pp r}-r^2\frac{\pp\Lambda}{\pp r}\frac{\pp\Phi}{\pp r}
+r^2\frac{\pp^2\Phi}{\pp r^2}+4r\frac{\pp\Phi}{\pp r}\right)f_R'' \non
&& +e^{\Phi-\Lambda}\mathcal{F}^{-2}\left(-6\frac{\pp\Lambda}{\pp r}-6r\frac{\pp\Lambda}{\pp r}\frac{\pp\Phi}{\pp r}
+6r\frac{\pp^2\Phi}{\pp r^2}+12\frac{\pp\Phi}{\pp r}+6r\left(\frac{\pp\Phi}{\pp r}\right)^2\right)f_R' \non
&& -e^{\Phi+\Lambda}r^2\mathcal{F}^{-2}\varrho''\nu^2\left[\frac{\pp\Lambda}{\pp r}\frac{\pp\Phi}{\pp r}
+2\left(\frac{\pp\Phi}{\pp r}\right)^2+\frac{4}{r}\frac{\pp\Phi}{\pp r}+\frac{\pp^2\Phi}{\pp r^2}\right] \non
&& -6re^{\Phi-\Lambda}\mathcal{F}^{-3}\left(\frac{3}{r}f_R'+f_R''\right)\frac{\pp\Phi}{\pp r}f_R'
+e^{\Phi+\Lambda}r^2\mathcal{F}^{-2}p\frac{\varrho'''\varrho'}{\varrho''}\nu^2\left(\frac{\pp\Phi}{\pp r}\right)^2 \non
&& +2e^{\Phi+\Lambda}r^2\mathcal{F}^{-3}\left(\frac{3}{r}f_R'+f_R''\right)
\varrho''\nu^2\frac{\pp\Phi}{\pp r}
+2e^{\Phi+\Lambda}r^2\mathcal{F}^{-2}\varrho'\nu\left(\frac{\pp\Phi}{\pp r}\right)^2
-e^{\Phi+\Lambda}r^2\frac{1}{2f_{RR}} \,.
\eean

Select all terms contain $\xi^2$ in \eq{delta2SFinal}, then the seventh term takes the form
\bean \label{C19}
\mathcal{P}_{Thermo}^7 \eqn \int_r 2e^{\Phi-\Lambda}\left[\left(2r\frac{\pp\Phi}{\pp r}+1\right)f_R
+\left(r^2\frac{\pp\Phi}{\pp r}+2r\right)f_R'\right]\mathcal{F}^{-2}\varrho'^2\nu^2e^{4\Lambda}\xi^2 \non
&& -e^{\Phi+\Lambda}r^2\frac{\varrho''}{\varrho'^2}\left(\varrho'\nu\frac{\pp\Phi}{\pp r}
+\frac{\varrho'^2}{\varrho''}\frac{\pp\Phi}{\pp r}-\frac{2}{r}\varrho'\nu-\varrho'\nu f_R''\mathcal{F}^{-1}\right)^2\xi^2 \non
&& -2e^{\Phi+\Lambda}r^2\varrho''\left(-\nu\frac{\pp\xi}{\pp r}\right)
\left(\nu\frac{\pp\Phi}{\pp r}+\frac{\varrho'}{\varrho''}\frac{\pp\Phi}{\pp r}
-\frac{2}{r}\nu-\nu f_R''\mathcal{F}^{-1}\right)\xi \,.
\eean
Simplifying \eq{C19} and using integration by parts, after some calculations we obtain that
\bean
&& \mathcal{P}_{Thermo}^7 \non
\eqn \int_r e^{\Lambda+\Phi}r^2\varrho'\nu
\left[e^{2\Lambda}\varrho'\nu\frac{3}{r}f_R'\mathcal{F}^{-2}+2\left(\frac{\pp\Phi}{\pp r}\right)^2+\frac{\pp\Phi}{\pp r}\frac{\pp\Lambda}{\pp r}
-\frac{1}{r}\frac{\pp\Phi}{\pp r}+\frac{1}{r}\frac{\pp\Lambda}{\pp r}-2f_R''\mathcal{F}^{-1}\frac{\pp\Phi}{\pp r}-\frac{1}{r}f_R''\mathcal{F}^{-1}-\frac{\pp^2\Phi}{\pp r^2} \right]\xi^2 \non
&& +e^{\Lambda+\Phi}r^2\varrho''\nu^2\left[-\left(\frac{\pp\Phi}{\pp r}\right)^2-2\frac{\pp\Phi}{\pp r}\frac{\pp\Lambda}{\pp r}
+\frac{6}{r}\frac{\pp\Phi}{\pp r}-\frac{1}{r^2}
-\frac{e^{2\Lambda}}{r^2}+e^{2\Lambda}\frac{R}{2}+3f_R''\mathcal{F}^{-1}\frac{\pp\Phi}{\pp r}
+f_R''\mathcal{F}^{-1}\frac{\pp\Lambda}{\pp r} \right. \non
&& \left. +f_R'''\mathcal{F}^{-1}-2f_R''^2\mathcal{F}^{-2}-\frac{2}{r^2}f_R''f_R\mathcal{F}^{-2}
-\frac{4}{r}f_R''f_R'\mathcal{F}^{-2} \right]\xi^2 \non
&& +e^{\Lambda+\Phi}r^2\frac{\varrho'''\varrho'}{\varrho''}\nu^2\left[\left(\frac{\pp\Phi}{\pp r}\right)^2-\frac{2}{r}\frac{\pp\Phi}{\pp r}
-\frac{\pp\Phi}{\pp r}f_R''\mathcal{F}^{-1}\right]\xi^2 \,,
\eean
hence
\bean \label{STherC7}
\mathcal{C}_{Thermo}^7 \eqn e^{\Lambda+\Phi}r^2\varrho'\nu
\left[e^{2\Lambda}\varrho'\nu\frac{3}{r}f_R'\mathcal{F}^{-2}+2\left(\frac{\pp\Phi}{\pp r}\right)^2+\frac{\pp\Phi}{\pp r}\frac{\pp\Lambda}{\pp r}-\frac{1}{r}\frac{\pp\Phi}{\pp r}
+\frac{1}{r}\frac{\pp\Lambda}{\pp r}-2f_R''\mathcal{F}^{-1}\frac{\pp\Phi}{\pp r}-\frac{1}{r}f_R''\mathcal{F}^{-1}-\frac{\pp^2\Phi}{\pp r^2} \right] \non
&& +e^{\Lambda+\Phi}r^2\varrho''\nu^2\left[-\left(\frac{\pp\Phi}{\pp r}\right)^2-2\frac{\pp\Phi}{\pp r}\frac{\pp\Lambda}{\pp r}
+\frac{6}{r}\frac{\pp\Phi}{\pp r}-\frac{1}{r^2}-\frac{e^{2\Lambda}}{r^2}+e^{2\Lambda}\frac{R}{2}+3f_R''\mathcal{F}^{-1}\frac{\pp\Phi}{\pp r}
 \right. \non
&& \left. +f_R''\mathcal{F}^{-1}\frac{\pp\Lambda}{\pp r} +f_R'''\mathcal{F}^{-1}-2f_R''^2\mathcal{F}^{-2}-\frac{2}{r^2}f_R''f_R\mathcal{F}^{-2}
-\frac{4}{r}f_R''f_R'\mathcal{F}^{-2} \right] \non
&& +e^{\Lambda+\Phi}r^2\frac{\varrho'''\varrho'}{\varrho''}\nu^2\left[\left(\frac{\pp\Phi}{\pp r}\right)^2-\frac{2}{r}\frac{\pp\Phi}{\pp r}
-\frac{\pp\Phi}{\pp r}f_R''\mathcal{F}^{-1}\right] \,,
\eean

The last term is the $b\xi$ term, this term is not complicated, so $\mathcal{C}_{Thermo}^8$ can be directly read off from \eq{delta2SFinal},
\bean \label{STherC8}
\mathcal{C}_{Thermo}^8 \eqn 2e^{\Phi-\Lambda}\left[-2r\frac{\pp\Lambda}{\pp r}
-r^2\frac{\pp\Lambda}{\pp r}\frac{\pp\Phi}{\pp r}
+r^2\left(\frac{\pp\Phi}{\pp r}\right)^2+r^2\frac{\pp^2\Phi}{\pp r^2}\right]
\mathcal{F}^{-1}(-\varrho'\nu e^{2\Lambda}) \non
&& +2e^{\Phi-\Lambda}\left[\left(2r\frac{\pp\Phi}{\pp r}+1\right)f_R
+\left(r^2\frac{\pp\Phi}{\pp r}+2r\right)f_R'\right]
\mathcal{F}^{-2}\left(2\varrho'\nu e^{2\Lambda}\frac{\pp\Phi}{\pp r} \right) \non
&& -2e^{\Phi+\Lambda}r^2\varrho''
\left(\nu\frac{\pp\Phi}{\pp r}+\frac{\varrho'}{\varrho''}\frac{\pp\Phi}{\pp r}
-\frac{2}{r}\nu-\nu f_R''\mathcal{F}^{-1}\right)\mathcal{F}^{-1}\nu\frac{\pp\Phi}{\pp r} \non
\eqn \mathcal{F}^{-1}\varrho'\nu e^{\Phi+\Lambda}\left(4r\frac{\pp\Lambda}{\pp r}+2r^2\frac{\pp\Lambda}{\pp r}\frac{\pp\Phi}{\pp r}
+2r\frac{\pp\Phi}{\pp r}-2r^2\frac{\pp^2\Phi}{\pp r^2}\right) \non
&& +6\mathcal{F}^{-2}\varrho'\nu e^{\Phi+\Lambda}\frac{\pp\Phi}{\pp r}r f_R'
-2\mathcal{F}^{-1}e^{\Phi+\Lambda}r^2\varrho''\nu^2\frac{\pp\Phi}{\pp r}
\left(\frac{\pp\Phi}{\pp r}-\frac{2}{r}-f_R''\mathcal{F}^{-1}\right) \,. \non
\eean

Substituting these coefficients $\mathcal{C}_{Thermo}^i$ into Eq. (\ref{ThermoResult}) we obtain the thermodynamical stability criterion for perfect fluid in $f(R)$ theories.
\end{widetext}

\end{document}